\documentclass[preprint2]{aastex631}

\usepackage{amsmath}

\RequirePackage{threeparttable}
\RequirePackage{booktabs}
\RequirePackage{multirow}
\usepackage{tabularx}
\newcolumntype{C}{>{\centering\arraybackslash}X}
\newcolumntype{L}{>{\raggedright\arraybackslash}X}

\usepackage{float}
\usepackage[titletoc]{appendix}
\usepackage{comment}

\begin{document}

\title{Effective Viscosity in the Intracluster Medium During Magnetic Field Amplification via Turbulent Dynamo}


\author[0000-0002-6374-9321]{S. Adduci Faria}
\affiliation{Instituto de Astronomia, Geofísica, e Ciências Atmosféricas - Universidade de São Paulo, \\
Cidade Universitária, R. do Matão, 1226, CEP 05508-090, São Paulo, SP, Brazil}

\author[0000-0001-6880-4468]{R. Santos-Lima }
\affiliation{Instituto de Astronomia, Geofísica, e Ciências Atmosféricas - Universidade de São Paulo, \\
Cidade Universitária, R. do Matão, 1226, CEP 05508-090, São Paulo, SP, Brazil}


\author[0000-0001-8058-4752]{E. M. de Gouveia Dal Pino}
\affiliation{Instituto de Astronomia, Geofísica, e Ciências Atmosféricas - Universidade de São Paulo, \\
Cidade Universitária, R. do Matão, 1226, CEP 05508-090, São Paulo, SP, Brazil}



\begin{abstract}

Galaxy clusters, the largest gravitationally bound structures, host a hot, diffuse plasma with poorly understood viscosity and magnetic field amplification.  
Astrophysical plasmas are often modeled with magnetohydrodynamics (MHD), but low collision rates in environments like the intracluster medium (ICM) hinder thermodynamic equilibrium, causing pressure anisotropies and high viscosity. High-\(\beta\) plasmas, dominated by thermal pressure, are prone to instabilities (e.g., firehose, mirror) that limit anisotropy, reduce viscosity, and enable small-scale dynamo-driven magnetic amplification.  
This study examines viscosity evolution in the ICM during turbulent magnetic field amplification. We performed 3D MHD simulations of forced turbulence with an initially weak, uniform magnetic field. Using the CGL-MHD framework, we incorporate anisotropic pressure dynamics and instability-driven anisotropy limitation. We analyze effective viscosity and dynamo evolution, comparing results with Braginskii-MHD and uniform-viscosity MHD.  
Results show viscosity decreases over time, allowing magnetic field amplification to saturation levels similar to non-viscous MHD. Viscosity distribution becomes bimodal, reflecting (i) collisional values and (ii) turbulence-dominated values proportional to $1 \times 10^{-4} L_{\rm turb} U_{\rm turb}$ in unstable regions. At saturation, 60\% of plasma retains collisional viscosity.  
Braginskii-MHD reproduces similar magnetic amplification and viscosity structures. However, uniform-viscosity MHD, where viscosity equals the mean saturated CGL-MHD value, fails to capture the turbulence inertial range. These findings highlight the need for anisotropic viscosity models in studying ICM processes like magnetic topology, cosmic ray transport, and AGN-driven shocks.  
Moreover, our CGL-MHD and Braginskii-MHD models match the Coma cluster density fluctuation spectrum, reinforcing its weakly collisional nature.

\end{abstract}


\section{Introduction} \label{sec:intro}

Clusters of galaxies are structures consisting of hundreds to thousands of gravitationally bound galaxies, with typical masses ranging from $10^{12}$ to $10^{15}$ solar masses, most of which are dark matter. The visible baryonic mass is mainly in the hot gas emitting X-rays that fills the intracluster medium (ICM). Galaxy clusters may contain valuable information about magnetic fields in the early Universe (\citealt{beresnyak2018nonlinear}).  Magnetic fields in these environments can be mainly inferred from the observation of synchrotron radiation in the radio band, which arises from relativistic electrons spiraling around magnetic field lines. Estimates based on synchrotron emission suggest field strengths ranging from $\sim 1 \mu$G at the center
\citep{willson1970radio, brunetti2001particle, grasso2001magnetic, kulsrud2008origin, guidetti2008intracluster, vallee2011magnetic, ryu2012magnetic, vacca2012intracluster, brunetti2014cosmic, govoni2017sardinia, bonafede2018constraining, stuardi2021intracluster}, to $ 0.1 - 0.3 \mu$G at the outskirts of the clusters \citep{kim1989discovery, stuardi2021intracluster}. Additionally, it is worth noting that galaxy clusters are interconnected by cosmic filaments, which exhibit magnetic field strengths typically below $0.25$ $\mu$G \citep{locatelli2021new}.

Unraveling the origin of these large-scale magnetic fields observed in galaxy clusters poses a significant challenge. Their sources remain elusive. One plausible explanation involves the folding and stretching of weak primordial magnetic fields through turbulent dynamo processes \citep{ryu2012magnetic, subramanian2016origin},
potentially accounting for the observed magnetic fields within clusters \citep[e.g. ][]{dolag1999sph, bruggen2005simulations, de2013turbulence, santos2014magnetic, vazza2018resolved}.
According to \citealt[][]{santos2014magnetic} and references therein, the turbulent dynamo transfers about 6\% of the energy flux of the hydrodynamic cascade into magnetic energy, and as the magnetic field enters into equipartition with hydrodynamic movements, it acquires correlation lengths of the order of the largest turbulence scales. This amplification process is independent of the value of the initial seed field and should be efficient (\citealt{cho2009growth, beresnyak2012universal}).  
However, astrophysical sources such as supernovae and active galactic nuclei (AGN) could also substantially contribute substantial to magnetization, especially in halos \citep[e.g., ][]{bertone2006magnetic, donnert2009cluster, marinacci2015large}.
Moreover, these systems, along with cluster mergers and accretion, serve as primary drivers of turbulence in the ICM.

While turbulent dynamo amplification complicates the distinction between these alternative scenarios for magnetic field origin within halos \citep{cho2014origin, beresnyak2016turbulent, vazza2018resolved},
more extended regions of the cosmic web are expected to undergo gentler evolution, retaining a dynamic memory of the origin of large-scale magnetic fields \citep{vazza2014amplification}. The combination of cosmological simulations exploring scenarios with primordial seed fields and observations assessing the statistical signature of diffuse primordial magnetic fields suggests the presence of very large-scale magnetic fields in the sparser regions of the Universe.  Gamma-ray observations of distant blazars also endorse these conclusions \citep[e.g., ][]{foffano2019extreme, CTAOcoll_2021, aschersleben2022cta}.
This phenomenon is possibly only explicable by primordial seeding scenarios \citep[e.g., ][]{dolag2005constrained, vazza2021simulations, carretti2023magnetic, vernstrom2023polarized}
subject to turbulent dynamo amplification.

The growth of the turbulent dynamo field in global galaxy cluster simulations has been mainly investigated through 3D collisional  MHD simulations with uniform viscosity by several authors \citep[e.g. ][]{dolag2005constrained,vazza2018resolved}. 
\citealt{vazza2018resolved}, in particular, performing high resolution simulations of a $\sim 10^{15} M_\odot$ cluster, obtained
magnetic fields (originating from primordial seeds with comoving intensities of $0.1$nG at redshift $z=30$), with values $\sim 2$ $\mu$G in the innermost region of the cluster which is compatible with observations and with the earlier cosmological simulations by, e.g., \citet[][]{dolag2005constrained, donnert2018magnetic, vazza2018resolved, dominguez2019dynamical}.

However, this scenario faces some fundamental difficulties. The typical parameters of the cluster environment  such as the ion number density $n_i \sim 10^{-3}$ cm$^{-3}$ and the temperature $T\sim 10 ^8$ K \citep{lagana2008star, zhuravleva2019suppressed}, yield collision rates for ions 
$\nu_{ii} \approx 317$ Myr$^{-1} \left(\frac{n_i}{10^{-3}\,cm^{-3}}\right) \left(\frac{10\,keV} {T_i}\right)^{\frac{3}{2}}$, 
and  mean free paths $\lambda_{ii}  \approx  100 \, {\rm kpc} \left(\frac{n_i}{10^{-3}\,cm^{-3}}\right) \left(\frac{10\,keV}{T_i}\right)^{2}$ \citep[][and references therein]{donnert2018magnetic}. 
These low-frequency collisions imply a predominantly non-collisional flow on the largest scales of the systems.
This leads to an anisotropy in the thermal motion of the particles between directions parallel and perpendicular to the magnetic field (\citealt{schekochihin2005plasma, kulsrud2005plasma}).
These departures from pressure isotropy apply anisotropic viscous stresses to the plasma motions, that may restrict their ability to elongate the magnetic field lines \citep{sharma2006shearing, santos2014magnetic, st2020fluctuation}. On the other hand, the same anisotropies trigger plasma instabilities, such as firehose and mirror \citep{kulsrud1983mhd, gary1993theory, gary2000electromagnetic, kunz2014firehose}, that may lead to an effective scattering of the ions, thereby allowing for a limitation at the anisotropy of their thermal motions, and the operation of the dynamo, through random shear.
Recent studies \citep{rosin2011non, santos2014magnetic, santos2016limits, santos2017features, kunz2014firehose, st2020fluctuation, squire2023pressure} have emphasized the need for a more refined approach that takes these phenomena into account for an accurate description of the environment's dynamics and, in special, of the turbulent dynamo action in the ICM.

Earlier works brought important results in modeling and understanding the growth of magnetic fields through the action of small-scale dynamos employing a collisionless MHD approach. The Chew-Goldberger-Low (CGL-MHD) model \citep{chew1956boltzmann} introduced the simplest approximation within this scenario, where the anisotropic components of the parallel ($p_\parallel$) and perpendicular ($p_\perp$) pressures to the magnetic field are constrained by the double-adiabatic closure. The 3D MHD simulations of turbulent flows with fixed pressure anisotropy by \citet{kowal2011turbulence} have demonstrated how the presence of instabilities caused by temperature anisotropy significantly alters the structure of the density, magnetic field, and velocity distributions of the gas. Moreover, they revealed a substantial increase in turbulent energy at small scales. 
 
On the other hand, \citet{santos2014magnetic} employed a modified CGL-MHD model where, as in the solar wind and magnetosphere's modeling, it was assumed that the growth of the kinetic instabilities constrains the anisotropy close to the threshold for these same instabilities. Under this approach, they performed 3D  simulations of the transonic, high $\beta$  turbulent ICM  (where $\beta$ is the ratio between the thermal and magnetic pressures) to investigate the amplification of magnetic fields. They found that with a relaxation rate of the anisotropy to the threshold values of the order of the growth rate of the kinetic instabilities, the magnetic seeds quickly amplify to values as large as those found in collisional MHD models. 

Another class of weak collisional models has employed an anisotropic pressure decomposition following  \citet{braginski1965transport} formalism \citep[e.g.][]{schnack2009lectures, devlen2010finite, st2020fluctuation}. 
\citet{st2020fluctuation}, in particular, including a field-parallel viscous (Braginskii) stress in the MHD equations, obtained results equivalent to those of \citet{santos2014magnetic}. As in that work, they found that when the stress is constrained within values consistent with a pressure anisotropy regulated by firehose and mirror instabilities, the Braginskii-MHD dynamo closely resembles its MHD counterpart. Conversely, when the parallel viscous stress is allowed to operate without the constraints (i.e. without the pressure-anisotropy limiters), it will not sustain a dynamo if the ratio of perpendicular to parallel viscosities is too small similarly to \citet{santos2014magnetic}.

Recent observations of the Coma cluster by the Chandra satellite \citep{zhuravleva2019suppressed}, have revealed that its turbulence cascades down to scales that are much smaller than those predicted by standard collisional fluid theory, by more than an order of magnitude. This implies 
the presence of a much smaller viscosity than that given by Coulomb collisions of the weakly collisional plasma (the Spitzer viscosity). The suppression of viscosity at these scales is probably due to an effective enhancement of particle collisions with micro-fluctuations caused by plasma instabilities. These findings suggest the necessity of a collisionless MHD approach, as described above, to investigate the turbulent dynamo amplification of magnetic fields in galaxy clusters and surroundings.

The understanding of viscosity in the intracluster medium (ICM) is essential for linking microscopic processes, such as anisotropic transport and energy dissipation, to the macroscopic dynamics that shapes the evolution of galaxy clusters.  
Recent studies highlight the relevance of these processes in various astrophysical contexts. For example, the qualification of substructures resulting from mergers in evolutionary processes within galaxy clusters through X-ray observations \citep{andrade2012new}.  
Studies on "jellyfish" galaxies, such as in the A901/2 system simulation \citep{ruggiero2019galaxy}, show that regions experiencing a sharp increase in ram pressure exhibit density and velocity gradients, which may be directly related to the effective viscosity of the ICM.  
In the Perseus cluster, turbulence generated by supernovae (SNe) and the dynamics of the active galactic nucleus (AGN) promotes the formation of magnetic filaments and loops \citep{Falceta-Goncalves2010a, falceta2010precessing}. These processes, influenced by plasma viscosity, are crucial for suppressing cooling flows and redistributing thermal and magnetic energy.  
The propagation of cosmic rays in the ICM is determined by the topology of magnetic fields, which is shaped by the small-scale turbulent dynamo process, where viscosity can have a direct influence. Organized structures create anisotropies in particle flux, as demonstrated in simulations of electromagnetic cascades and angular distributions around high-energy sources \citep{giacalone1999transport, sigl2004ultrahigh, alves2016probing}. Even weak magnetic fields ($\sim$nG) can cause significant deflections at ultra-high energies \citep{alves2017implications, hussainetal2023}. The intensity and coherence scale of these fields modulate diffusive behavior: while weak fields suppress flux less effectively, stronger and more organized fields enhance scattering and deflections \citep{batista2014diffusion, Alves-Batista_etal2019, hussain2021high, hussainetal2023}.

In this work, we investigate the evolution and structure of viscosity in the intra-cluster medium (ICM) during the turbulent amplification of magnetic fields through small-scale dynamos, considering MHD approaches that include the effects of weakly collisional plasma. 
Section \ref{sec:02} provides an overview of the weakly collisional MHD models investigated in this study. Section \ref{sec:03} outlines the numerical methods and simulation setup employed. Section \ref{sec:04} presents the results of our analysis, followed by a discussion in Section \ref{sec:05}. Finally, Section \ref{sec:06} concludes the work by summarizing the main findings.

\section{Single fluid models for the ICM}\label{sec:02}

In this Section, we 
introduce our
primary model (Model A, \S~\ref{sec:cgl_mhd}), which 
describes the weakly collisional ICM plasma as a single  MHD fluid with anisotropic pressure. This model 
extends the CGL double-adiabatic closure to incorporate
a parameterized non-adiabatic evolution of the pressure anisotropy, capturing microphysical processes in the plasma.
This model does not use an explicit viscous term in the equations, but 
we will demonstrate that an effective viscosity coefficient can be inferred 
by interpreting the anisotropic pressure stress as a viscous term contribution.
Additionally, 
for comparison, we consider simpler MHD models previously employed in ICM studies: 
Model B (\S~\ref{sec:braginskii_mhd}),  
which corresponds to standard MHD with an anisotropic viscous stress tensor (Braginskii viscosity), and Model MHD (\S~\ref{sec:isotropic_mhd}), representing the standard collisional MHD with isotropic viscosity.

\subsection{Model A: Extended CGL-MHD} \label{sec:cgl_mhd}

As stressed in \S~\ref{sec:intro}, over the scales of the plasma characterized by a relatively 
low rate of coulomb collisions between particles, anisotropies in the thermal pressure can emerge. In a single fluid description (MHD), this anisotropy can be translated in different temperatures (or pressures), parallel and perpendicular to the magnetic field, reflecting distinct dispersion in the velocity distributions of particles 
along these directions.

Following \citet{santos2014magnetic}, the anisotropic MHD equations can be formulated in the 
following way:
\begin{equation*}
  \frac{\partial }{\partial t}
  \begin{bmatrix}
    \rho \\[6pt]
    \rho \mathbf{u} \\[6pt]
    \mathbf{B} \\[6pt]
    A (\rho^{3}/B^{3})
  \end{bmatrix}
  + \nabla \cdot
  \begin{bmatrix}
    \rho \mathbf{u} \\[6pt]
    \rho \mathbf{uu} + p\mathbf{I} + \Pi_{\rm M} \\[6pt]
    \mathbf{uB - Bu} \\[6pt]
    A (\rho^{3}/B^{3}) \mathbf{u}
  \end{bmatrix}
  =
\end{equation*}
\begin{equation}
  = 
  \begin{bmatrix}
    0 \\[6pt]
    \nabla \cdot \Pi_{\delta p} + \mathbf{f} \\[6pt]
    0 \\[6pt]
    \dot{A}_{S} (\rho^{3}/B^{3})
  \end{bmatrix}
  \rm{,}
\label{eqn:collisionless_mhd}
\end{equation}
where $\rho$ and $\mathbf{u}$ are the fluid density and velocity fields, respectively, 
$A = p_{\perp} / p_{\parallel}$ is the ratio between thermal pressures 
perpendicular and parallel to the local magnetic field $\mathbf{B}$, $\dot{A}_{S}$ 
is the rate of change of the local anisotropy due to microphysical plasma processes,
$\mathbf{f}$ represents any external bulk force (in our study this force is responsible 
for driving the turbulence in the system), 
and the tensors $\Pi_{\delta p}$ and $\Pi_M$ are defined by:
\begin{equation}\label{eq:viscosidade_efetiva}
\Pi_{\delta p} = - \delta p \left( \frac{1}{3} \mathbf{I} - \mathbf{bb} \right)\rm{,}
\end{equation}
and
\begin{equation}
\Pi_{\rm M} = (B^{2}/8 \pi) \mathbf{I} - \mathbf{BB} /4 \pi \rm{,}
\end{equation}
where $\mathbf{I}$ is the unitary dyadic tensor, $\mathbf{b} = \mathbf{B} / B$, 
and $\delta p \equiv (p_{\perp} - p_{\parallel})$. 

In \citet{santos2014magnetic}, the system above is completed by an equation that evolves the 
internal energy, including thermal relaxation due to 
heat conduction and radiative cooling. 
Due to the low density of the ICM plasma, we assume that the environment is dominated by an optically thin gas, where radiative losses occur in equal proportion to heating very quickly, thus leading to an isothermal equation of state (EOS):
\begin{equation}\label{eq:isothermal}
p = \left(\frac{1}{3} p_{\parallel} + \frac{2}{3} p_{\perp} \right) = \rho c_s^2 \rm{,}
\end{equation}
where $c_s$ is an effective thermal speed. With this isothermal prescription, we represent an 
idealized system which rapidly (compared to the dynamical timescales of interest in our study) relaxes the total temperature to a specific value, but without changing the anisotropy $A$ during this relaxation process.

The equation which describes the evolution of $A$ (fourth row in equation~\ref{eqn:collisionless_mhd}), in the absence of source terms, represents the CGL (or double-adiabatic) closure \citep{chew1956boltzmann}, which neglects any parallel heat conduction and Landau damping effects that can be significant in the collisionless regime. It conserves the first adiabatic invariant for charged particles and the total entropy of the gas ($\propto p_{\perp}^{2/3} p_{\parallel}^{1/3} / \rho^{5/3}$), and is usually expressed in terms of the following equations:
\begin{equation}\label{eq:CGL}
    \frac{d}{dt} \left( \frac{p_\perp}{\rho B} \right) = 0, \hspace{1cm} \frac{d}{dt} \left( \frac{p_\parallel B^2}{\rho^3} \right) = 0.
\end{equation}

As in \citet{santos2014magnetic}, the CGL closure is modified to include anisotropy relaxation effects due to microphysical processes, which are represented by the term $\dot{A}_{S}$ in equation~\ref{eqn:collisionless_mhd}. These processes break the first adiabatic invariant, increasing the entropy of the gas, and therefore are irreversible. As examples of these microphysical processes, we can cite the coulomb collision between particles and the resonant scattering of particles by plasma waves (see e.g. \citealt{santos2016limits}). We shall represent generically these effects by an effective scattering rate $\nu_S$ which reduces the difference in the pressure components $\delta p \equiv (p_{\perp} - p_{\parallel})$ according to the relation
\begin{equation} \label{eq:deltap_evol}
\left( \frac{\partial \delta p}{\partial t} \right)_{\rm S} = - \nu_{\rm S} \delta p,
\end{equation}
or, translating to the rate of change of the pressure anisotropy field:
\begin{equation}\label{eq:dadt_nu}
\dot{A}_{S} \equiv \left( \frac{\partial A}{\partial t} \right)_{\rm S} = - \frac{ \nu_{\rm S} }{3} \left( 2A^2 - A - 1 \right).
\end{equation}

In  Model A, we assume that  the unperturbed plasma holds a scattering rate given by 
$\nu_{S} =  \nu_{S,0} \frac{\rho}{\rho_0}$, where $\rho_0$ is the unperturbed density, and $\nu_{S,0}$ is a fixed value, determined by 
 Coulomb collision between particles. 
During the system evolution, $A$ can attain values that trigger the excitation of unstable plasma waves, resulting in additional scattering due to wave-particle interactions (described in \S~\ref{sec:limits_anisotropy}). These interactions prevent $A$ from decreasing or increasing  significantly beyond the instability threshold.
In such situations, we estimate the new local scattering rate $\nu_{S}$ according to equation~\ref{eq:dadt_nu}:
\begin{equation} \label{eq:nu_scatt_instab}
	\nu_{\rm S} = - \frac{ 3 }{ \left( 2A^2 - A - 1 \right) } \left( \frac{\partial A}{\partial t} \right)_{\rm S} \rm{,}
\end{equation}
where, to keep $A$ 
within the threshold, 
\begin{equation} \label{eq:dadt_scatt_instab}
\left( \frac{\partial A}{\partial t} \right)_{\rm S} = 
- \left( \frac{\partial A}{\partial t} \right)_{\rm CGL} = 
\nabla \cdot \left(A \mathbf{u} \right) - 3 A \mathbf{b} \cdot \left[ \left( \mathbf{b} \cdot \nabla \right) \mathbf{u} \right] \rm{.}
\end{equation}

In analogy to the Braginskii viscosity of Model B, described in \S~\ref{sec:braginskii_mhd}, we define and quantify the effective kinematic viscosity coefficient for our Model A as 
\begin{equation} \label{eq:eta_eff}
\eta_{\rm eff} \equiv \frac{1}{\rho} \frac{p_{\parallel}}{\nu_{\rm S}}  \rm{,}
\end{equation}
which can also be used to evaluate the Reynolds number of the system.

\subsection{Model B: Braginskii-MHD} \label{sec:braginskii_mhd}

Under some simplifying hypotheses, we can reduce the set of equations~\ref{eqn:collisionless_mhd} by replacing the last dynamical equation with a simple formula for the instantaneous local value of $A$. This also leads to a description of the anisotropic part of the stress tensor, $\Pi_{\delta p}$ (Eq.~\ref{eq:viscosidade_efetiva}), as an effective viscous stress tensor in the momentum equation (Braginskii viscosity), at the same time allowing one to work with a single scalar pressure field $p$ (Eq.~\ref{eq:isothermal}).
Below we describe how we can connect this formulation with the previous one.

First we re-write the last equation of~\ref{eqn:collisionless_mhd} in the 
Lagrangean form
\begin{equation}
\frac{d A}{d t} = - A \nabla \cdot \mathbf{u} + 3 A \mathbf{b} \cdot \left[ \left( \mathbf{b} \cdot \nabla \right) \mathbf{u} \right] - \frac{ \nu_{\rm S} }{3} \left( 2A^2 - A - 1 \right).
\end{equation}

Observe that $A = 1 + \delta p / p_{\parallel}$. If 
we assume the pressure anisotropy to be close to unity over the dynamical timescales of interest (for example due to a short timescale of relaxation $\nu_S^{-1}$), then $\delta p / p_{\parallel} \ll 1$. In this sense, 
assuming $dA/dt \approx 0$,
\begin{equation}
	- A \nabla \cdot \mathbf{u}
	+ 3 A \mathbf{b} \cdot \left[ \left( \mathbf{b} \cdot \nabla \right) \mathbf{u} \right]
	- \frac{ \nu_{\rm S} }{3} \left( 2A^2 - A - 1 \right) \approx 0 \rm{.}
\end{equation}
Keeping only terms up to first order in $\delta p / p_{\parallel}$,
\begin{equation}
	- \nabla \cdot \mathbf{u}
	+ 3 \mathbf{b} \cdot \left[ \left( \mathbf{b} \cdot \nabla \right) \mathbf{u} \right]
	- (\delta p / p_{\parallel}) \nu_{\rm S} \approx 0 \rm{.}
\end{equation}

Therefore, 
\begin{equation} \label{eq:delta_p_braginskii}
	\delta p \approx \left( \frac{p_{\parallel}}{\nu_{\rm S}} \right) \left\{ - \nabla \cdot \mathbf{u}
	+ 3 \mathbf{b} \cdot \left[ \left( \mathbf{b} \cdot \nabla \right) \mathbf{u} \right] \right\} \rm{,}
\end{equation}
and now we have a closure for $\delta p$ which gives its instantaneous values according to the 
fields $p_{\parallel}, \mathbf{b, u}$ and the rate $\nu_S$.

Defining \footnote{Note that for an initial unperturbed system, $\eta_B$ holds the same value of  $\eta_{\rm{eff}}$ (eq. \ref{eq:eta_eff}) of model A.}
\begin{equation}\label{eq:eta_braginskii}
\eta_{\rm B} \equiv \frac{1}{\rho} \frac{p_{\parallel}}{\nu_{\rm S}}
\end{equation}
as a kinematic viscosity coefficient,
the anisotropic part of the stress tensor ($\Pi_{\delta p}$ on the rhs of Eq.~\ref{eq:viscosidade_efetiva}) takes the form of the Braginskii parallel viscous stress
\begin{equation}\label{eq:plasma_equation_braginskii}
\Pi_{\rm B} = - \rho \eta_{\rm B} \left\{ - \nabla \cdot \mathbf{u}
	+ 3 \mathbf{b} \cdot \left[ \left( \mathbf{b} \cdot \nabla \right) \mathbf{u} \right] \right\}
        \left( \frac{1}{3} \mathbf{I} - \mathbf{b b} \right) \rm{,}
\end{equation}
which is the form of the stress tensor of the  magnetized plasma (up to first order in $1/(\omega \tau)$, where $\omega$ and $\tau$ are the gyrofrequencies and the effective collision time for the ions, 
respectively).

We can therefore write down the Braginskii-MHD set of equations as:
\begin{equation*}
  \frac{\partial }{\partial t}
  \begin{bmatrix}
    \rho \\[6pt]
    \rho \mathbf{u} \\[6pt]
    \mathbf{B} \\[6pt]
  \end{bmatrix}
  + \nabla \cdot
  \begin{bmatrix}
    \rho \mathbf{u} \\[6pt]
    \rho \mathbf{uu} + p\mathbf{I} + \Pi_{\rm M} \\[6pt]
    \mathbf{uB - Bu} \\[6pt]
  \end{bmatrix}
  =
\end{equation*}
\begin{equation}
  = 
  \begin{bmatrix}
    0 \\[6pt]
    \nabla \cdot \Pi_{\rm B} + \mathbf{f} \\[6pt]
    0 \\[6pt]
  \end{bmatrix}
  \rm{.}
\label{eqn:braginskii_mhd}
\end{equation}
This system is also closed by the isothermal equation of state (Eq.~\ref{eq:isothermal}). 
A similar set of Braginskii-MHD equations was used in the work by 
\citet{st2020fluctuation}, where the authors investigated the impact of this approach on the regulation and behavior of the dynamo within the magnetized kinetic regime, although in the incompressible limit. Unlike the Model A, the Braginskii-MHD model requires a non-zero 
effective scattering rate (either collisional or due to wave-particle scattering)
for prescribing the 
pressure difference $\delta p$ (Eq.~\ref{eq:delta_p_braginskii}).

We consider in this case the unperturbed plasma to hold a fixed kinematic viscosity coefficient $\eta_{B,0}$ due to the Coulomb collision of particles. However, similar to Model A, 
when the local pressures difference 
$\delta p$ (or the anisotropy $A = 1 + \delta p / p_{\parallel})$ achieves values 
for which unstable plasma waves are excited and provide additional scattering 
due to wave-particle interactions, we consider these waves to prevent additional 
modification of $\delta p$ beyond the threshold $\delta p_{\rm th}$ for the instability. In this situation, the local 
kinematic viscosity coefficient can be evaluated following the prescription for $\delta p$
(Eq.~\ref{eq:delta_p_braginskii}) together with the definition of $\eta_{\rm B}$ (Eq.~\ref{eq:eta_braginskii}):
\begin{equation}\label{eq:eta_B-eq18}
	\eta_{\rm B} = \frac{\delta p_{\rm th}}{\rho} \left\{ - \nabla \cdot \mathbf{u}
	+ 3 \mathbf{b} \cdot \left[ \left( \mathbf{b} \cdot \nabla \right) \mathbf{u} \right] \right\}^{-1} \rm{.}
\end{equation}

\subsection{Model MHDV: with isotropic pressure  and isotropic viscosity} \label{sec:isotropic_mhd}

We also consider the standard collisional 
MHD equations with an isotropic pressure and  isotropic viscosity tensor:
\begin{equation*}
  \frac{\partial }{\partial t}
  \begin{bmatrix}
    \rho \\[6pt]
    \rho \mathbf{u} \\[6pt]
    \mathbf{B} \\[6pt]
  \end{bmatrix}
  + \nabla \cdot
  \begin{bmatrix}
    \rho \mathbf{u} \\[6pt]
    \rho \mathbf{uu} + p\mathbf{I} + \Pi_{\rm M} \\[6pt]
    \mathbf{uB - Bu} \\[6pt]
  \end{bmatrix}
  =
\end{equation*}
\begin{equation}
  = 
  \begin{bmatrix}
    0 \\[6pt]
    \nabla \cdot \Pi_{\rm I} + \mathbf{f} \\[6pt]
    0 \\[6pt]
  \end{bmatrix}
  \rm{,}
\label{eqn:isotropic_mhd}
\end{equation}
where the components of the isotropic viscous stress tensor $\Pi_{\rm I}$ are given by
\begin{equation}
\Pi_{{\rm I}, \alpha \beta} = \rho \eta \left( \frac{\partial u_{\alpha}}{\partial x_\beta} + \frac{\partial u_{\beta}}{\partial x_{\alpha}} - \frac{2}{3} \delta_{\alpha \beta} \nabla \cdot \mathbf{u} \right) \rm{,}
\end{equation}
with $\eta$ being the kinematic viscosity coefficient. For this model, 
we consider $\eta$ as a fixed parameter. 

This system is also closed by the isothermal equation of state (Eq.~\ref{eq:isothermal}).

\subsection{Hard-wall anisotropy limits} \label{sec:limits_anisotropy}

As mentioned in \S~\ref{sec:cgl_mhd} for Model A and in \S~\ref{sec:braginskii_mhd} for Model B, 
when the pressure anisotropy $A$ deviates from unity, it can achieve values for which 
kinetic theory predicts the appearance of plasma waves due to specific kinetic instabilities. These plasma waves grow faster 
at the scales close to the particle's Larmor radius, and therefore are expected to scatter efficiently the particles 
through resonant wave-particle interactions, relaxing the anisotropy to values close to the threshold 
for the specific instability (see, e.g., \citealt{santos2016limits}). 
In \citet{santos2014magnetic}, the mirror and firehose instabilities were considered as the primary mechanisms for limiting the anisotropy $A$ \footnote{See however \citep{melville2016pressure} for the limitations of this scenario for the mirror instability.}.
According to the kinetic theory, the criteria for the appearance of mirror instability is given by:
\begin{equation}\label{eq:mirror}
    A > 1 + \beta_\perp^{-1},
\end{equation}
where $\beta_\perp = p_\perp/(B^2/8\pi)$. 

The criteria for the development of the firehose instability is:
\begin{equation}\label{eq:firehose}
    A < 1 - 2 \beta_\parallel^{-1} \rm{,}
\end{equation}
where $\beta_{\parallel} = p_{\parallel}/(B^2/8\pi)$.

In practice, the anisotropy $A$ in Models A and B is constrained by the thresholds of these instabilities, which are referred to as hard-wall limiters, which keep 
$A$ inside the limits:
\begin{equation}\label{eq:hard-wall}
    1 -\frac{2}{\beta_\parallel} < A < 1 + \frac{1}{\beta_\perp}.
\end{equation}
This aligns with the pioneering work 
by \cite{sharma2006shearing} and, since then, it has been used in several weakly collisional and collisionless MHD simulations (e.g. \citealt{santos2014magnetic, Nakwacki2016, santos2017features, st2020fluctuation}).

\section{Numerical Methods}\label{sec:03}

We have performed numerical simulations to reproduce local conditions in the ICM with forced turbulence using the MHD models described in Section~\ref{sec:02},  which have been implemented in the Godunov-type modular code PLUTO \citep{mignone2007pluto}. The set of eqs. \ref{eqn:collisionless_mhd},\ref{eqn:braginskii_mhd}, and \ref{eqn:isotropic_mhd}
were evolved in a three-dimensional (3D) Cartesian box using periodic boundary conditions, to emulate the large scales of the ICM. Temporal integration has been performed using the second-order Runge-Kutta method (RK2), while spatial integration used the piecewise parabolic method (PPM) and a Riemann solver HLL, where flux contributions are evaluated in all directions simultaneously. 

At each sub-step (predictor or corrector) of the numerical time integration, 
the contribution from the 
conservative fluxes and the source terms are calculated separately. We should point 
out that for Model A, the equation for the pressure anisotropy A (fourth row in Eq.~\ref{eqn:collisionless_mhd}) is numerically solved in the same formulation as in \citet{santos2014magnetic}, which in the absence of source terms reads:
\begin{equation}\label{eq:loga_conservative}
\frac{\partial}{\partial t} \left\{ \rho \ln \left( A \rho^2 / B^3 \right) \right\} +
\nabla \cdot \left\{ \rho \ln \left( A \rho^2 / B^3 \right) \mathbf{u} \right\} = 0 \rm{.}
\end{equation}
During each sub-step of the RK2 predictor-corrector schemes, after the evolution of $A$ taking into account only the conservative flux in the last equation, the effects of $\dot{A}_{S}$ are included in the following way. First, we recover the separated components $p_{\parallel, \perp}$, and evolve their difference $\delta p$ during the time interval using the analytical solution from Eq.~\ref{eq:deltap_evol}. If 
the final value of $A$ is beyond the hard-wall boundaries given by Eq.~\ref{eq:hard-wall} for the local value of $p$ and $\beta$, then $A$ is replaced by the corresponding threshold value $A_{\rm th} (p, \beta)$. The scattering rate due to the instability is calculated using Eqs.~\ref{eq:nu_scatt_instab} and~\ref{eq:dadt_scatt_instab} with $A = A_{\rm th}$ in the latter case.

In a similar way, during the calculation of the viscous stress tensor of 
Model B (Eq.~\ref{eq:plasma_equation_braginskii}), we test if the $\delta p= (A-1)p_{\parallel}$ obtained 
from eqs.~\ref{eq:delta_p_braginskii} and~\ref{eq:eta_braginskii} are 
inside the instability limiters (Eq.~\ref{eq:hard-wall}), otherwise it is replaced by  the corresponding threshold $\delta p_{\rm th}  (p, \beta)$.  
The local kinematic viscosity 
$\eta_B$, given by the enhanced scattering provided by the plasma waves, 
is then re-evaluated using Eq. \ref{eq:eta_B-eq18}.

Our models do not consider non-ideal dissipative effects in the induction equation, such as ambipolar diffusion or Hall effect, which are negligible in the scales considered. We also neglect thermal conduction.

\subsection{Code units} \label{sec:code_units}

The calculations are performed in code units. We have arbitrarily selected the initial density of the system ($\rho_0$), the length of the box (i.e. the length scale of the system $L$), and the isothermal sound speed ($c_s$) as unities in the code. The magnetic field in the code units absorbs the factor $1/\sqrt{4\pi}$, distinctly from the physical dimensions used in the previous equations (in other words, it is given in units of $\sqrt{\rho_0} c_s$).

To obtain the values of all quantities in physical units, we have adopted an initial density from observations $\rho_0 = 1.67 \times 10^{-27}$ g/cm$^3$ \citep[e.g.][]{chardin2018self}, a length of the box given by the typical size of a cluster $L=0.5$ Mpc \citep[e.g.][]{willson1970radio, brunetti2001particle}, and an isothermal sound speed $c_{s} = 2 \times 10^8$cm/s (\citealt{sarazin1986x}). 
Therefore, the magnetic field can be converted from code units $B_{\rm cu}$ to cgs physical units $B_{\rm cgs}$ through the relation $B_{\rm cgs} = \sqrt{4\pi\rho_0} c_{s} B_{\rm cu}$.

\subsection{Simulations setup and initial parameters}
\label{sec:setup}

\begin{table*}[!thbp]
\centering
\caption{Parameters of the Simulated Models.} 
\begin{tabular}{l c c c c c c c}
\toprule									
{Run} & 
{$ \dfrac{\eta_{0}^{*}}{\left(  L_{\rm turb} \;  U_{\rm turb} \right)}$} &
{$Re_0 \equiv L_{\rm turb} \dfrac{U_{\rm turb}}{\eta_{0}^{*}}$} &
{$\beta_{0} \equiv \dfrac{2 c_s^2}{v_{A,0}^2}$} & 
{$\mathcal{M}_{S, 0} \equiv \dfrac{U_{\rm turb}}{c_s}$} &
{$\mathcal{M}_{A, 0} \equiv \dfrac{U_{\rm turb}}{v_{A,0}}$} & 
{$\dfrac{t_{\rm final}}{(L_{\rm turb} / U_{\rm turb})}$} & 
{Res} \\
\midrule
\midrule
{A064HW} & 
{$0.0167$} &
{$30$} &
{$2 \times 10^{6}$} & 
{$1$} & 
{$10^{3}$} & {$40$} & 
{$64^{3}$} \\ [3pt]
\midrule
{A128HW} & 
{$0.0167$} &
{$30$} &
{$2 \times 10^{6}$} & 
{$1$} & 
{$10^{3}$} & 
{$40$} & 
{$128^{3}$} \\ [3pt]
\midrule
{A256HW} & 
{$0.0167$} &
{$30$} &
{$2 \times 10^{6}$} & 
{$1$} & 
{$10^{3}$} & 
{$40$} & 
{$256^{3}$} \\ [3pt]
\midrule
{A256HWnG} & 
{$0.0167$} &
{30} &
{$2 \times 10^{10}$} & 
{$1$} & 
{$10^5$} & 
{$40$} & 
{$256^{3}$} \\ [3pt]
\midrule
{B064HW} & 
{$0.0167$} &
{30} &
{$2 \times 10^{6}$} & 
{$1$} & 
{$10^{3}$} & 
{$40$} & 
{$64^{3}$} \\ [3pt]
\midrule
{B128HW} & 
{$0.0167$} &
{30} &
{$2 \times 10^{6}$} & 
{$1$} & 
{$10^{3}$} & 
{$40$} & 
{$128^{3}$} \\ [3pt]
\midrule
{B256HW} & 
{$0.0167$} &
{30} &
{$2 \times 10^{6}$} & 
{$1$} & 
{$10^{3}$} & 
{$40$} & 
{$256^{3}$} \\ [3pt]
\midrule
{MHD256} & 
{$0$} &
{$(\sim 1000)$} &
{$2 \times 10^{6}$} & 
{$1$} & 
{$10^{3}$} & 
{$40$} & 
{$256^{3}$} \\ [3pt]
\midrule
{MHDV256} & 
{$0.0167$} &
{$30$} &
{$2 \times 10^{6}$} & 
{$1$} & 
{$10^{3}$} & 
{$40$} & 
{$256^{3}$} \\ [3pt]
\bottomrule									
\end{tabular}
\label{tab:initial_condition}
\end{table*}

In all our simulations, we start the system at rest with homogeneous conditions. At each time step of the numerical integration, we introduce motions in the plasma on scales comparable to the computational domain size $L$. Operationally, we add velocity fields represented by Fourier discrete modes, with wavelengths $0.5 < \lambda / L \ < 1$ and a random phase distribution picked at each time step. This procedure creates motions with correlation scale $L_{\rm turb} \approx 0.5 L$. The mechanical power is kept constant throughout the simulation, and we use the same for all simulations. 
{\color{black}This assumption is energetically justified in the context of the intracluster medium (ICM), where turbulence power is thought to be roughly balanced by radiative losses from Bremsstrahlung emission \citep{mohapatra2019turbulence}.}
After the turbulence fully develops, the rms velocity of our MHD simulations without explicit viscosity, $U_{\rm turb}$, reaches values very close to the sound speed  $c_s$, also kept fixed in all our simulations. From this point forward, we define $U_{\rm turb}= c_s$ as the initial value of the system's turbulent velocity $V_{rms}$. Considering a turbulent hydrodynamic cascade, we can estimate the total power injected or dissipated into the system through the relation $P_{\rm diss} \sim \rho_0 U_{\rm turb}^3 / L_{\rm turb}$. Using the choice of physical units of \S~\ref{sec:code_units}, we have $P_{\rm diss} \sim  1.6 \times 10^{-26}$ erg cm$^{-3}$ s$^{-1}$.

In the initial conditions, we consider two values for the intensity of the seed magnetic field (which is uniform in our setup), with energy densities corresponding to the parameter $\beta_{0} \equiv \rho_0 c_s^2 / (B_{0}^2 / 8 \pi) = 2 \times 10^{6}$ and $2 \times 10^{10}$. Using physical units from \S~\ref{sec:code_units}, these values amount to an initial magnetic intensity $B_{0} = 1.44 \times 10^{-8}$ G and $1.44 \times 10^{-10}$ G, respectively. 
{\color{black} The goal is to follow their amplification by turbulence until the field reaches dynamically relevant strengths, consistent with observed values 
within a physically plausible timescale for the universe's evolution.}


The initial value of the kinematic viscosity for the different models, represented by $\eta_{0}^{*}$ (equivalent to $\eta_{{\rm eff}, 0}$, $\eta_{{\rm B}, 0}$, or $\eta_0$, for models A, B and MHDV, respectively) are chosen according to the estimated rate of coulomb ion-ion collisions,  $\eta_0^{*} = 0.0167 \; Lv_{s0}$. 
{\color{black}Using the physical units adopted in} \S~\ref{sec:code_units},
{\color{black}this corresponds to} $\eta_0^{*} = 5 \times 10^{30}$ cm$^{2}$ s$^{-1}$. 
In one of our simulations, {\color{black}which was} used for comparison, there is no explicit viscosity, i.e., $\eta_{0}^{*} = 0$. However, there is always a non-negligible numerical viscosity,  which we can estimate {\color{black}by comparing the velocity power spectra from turbulent simulations with varying levels of explicit viscosity. By identifying the viscosity value at which the spectral amplitude begins to diminish at smaller scales of the inertial range, we infer a numerical viscosity on the order of $\eta_{\rm num} \lesssim 10^{-3} L \, U_{\rm rms}$ for simulations with resolution $256^3$, assuming an injection scale $L_{\rm inj} \sim 0.5 \, L$.
Converting to physical units and using $U_{\rm rms} \sim c_s$, we find $\eta_{\rm num} \lesssim 3 \times 10^{29}$ cm$^{2}$ s$^{-1}$.}

We evolve the system until a time $t_{\rm final}=80 (L_{\rm turb} / U_{\rm turb})$, which in physical units (\S~\ref{sec:code_units}) is equivalent to $t_{\rm final} \approx 9.56$ Gyr.

Table~\ref{tab:initial_condition} presents the parameters of all our numerical simulations. In the first column, each run is labeled according to the plasma model: `A', `B', or `MHDV' (see \S~\ref{sec:cgl_mhd},~\ref{sec:braginskii_mhd}, and~\ref{sec:isotropic_mhd}, respectively). The `MHD' run (without the final `V') uses the Model MHD  without explicit kinematic viscosity. The number that follows stands for the resolution of the 3D simulation domain. In the runs using Model A and Model B, the names are also followed by `HW' to remind us of the presence of the anisotropy hard-wall limiters (\S~\ref{sec:limits_anisotropy}). The second column shows the initial collisional viscosity $\eta_{0}^{*}$ in units of $L c_{s}$, and the third column gives an estimate of the Reynolds number from the simulations considering this collisional viscosity $\eta_{0}^{*}$. Columns 4, 5, and 6 show the values of $\beta_0$, the initial sonic ($M_{S,0}$) and Alfvénic ($M_{A,0}$) Mach numbers, respectively, calculated using the initial values of the rms turbulence velocity $U_{\rm turb}$, the sound speed $c_s$,  the uniform density $\rho_0$ and magnetic field intensity $B_0$. Column 7 shows the final time of the simulations $t_{\rm final}$ in units of $L_{\rm turb} / U_{\rm turb}$, and the last column shows the simulation resolution. We note that all models share the same initial conditions, except for the resolution and the Reynolds number in the non-viscous MHD model. Since our primary goal here is to directly compare different model classes (A, B, MHDV, and MHD), we have maintained consistent fiducial parameter values whenever possible. Nevertheless, we include them in the table for clarity.

\section{Results}\label{sec:04}

\subsection{Amplification of Seed Magnetic Fields}

\begin{figure}
    \centering
    \includegraphics[width=1.0\columnwidth]{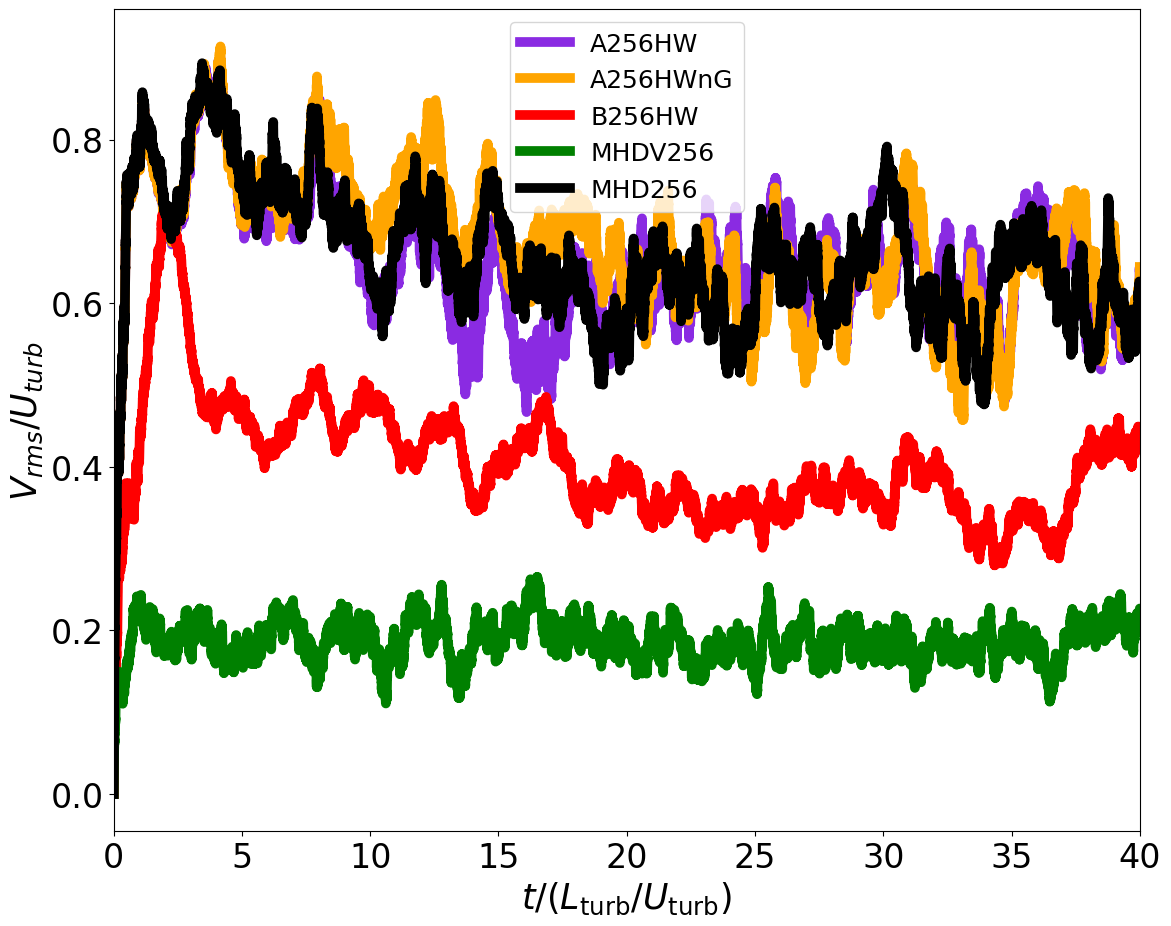}
    \caption{Temporal evolution of the $rms$ velocity, normalized by the initial value $U_{\rm turb}$, for models A256HW, A256HWnG, B256HW, MHDV256, and MHD256. Model parameters are listed in Table~\ref{tab:initial_condition}.}
    \label{fig:compared_graphicsV}
\end{figure}

Figure~\ref{fig:compared_graphicsV} shows the evolution of the turbulent $rms$ velocity in the domain for the runs listed in Table~\ref{tab:initial_condition} with a resolution of $256^3$. In all runs, turbulence is apparently fully developed from time $t \approx 5 \; (L_{\rm turb}/U_{\rm turb})$, with the $rms$ velocity remaining approximately constant until the end of the simulation. In run A256HW, the $rms$ velocity reaches around $6 \times 10^{-1} \; c_{s}$, while in run B256HW, this value is slightly lower, close to $4 \times 10^{-1} \; c_{s}$. The run MHDV256 stabilizes at approximately $2 \times 10^{-1} \; c_{s}$, reflecting a lower kinetic energy level than the other models. The run MHD256, which lacks anisotropic viscosity, displays an $rms$ velocity similar to runs A256HW and A256HWnG.

{\color{black} 
We note that although the
 injected power in the simulations corresponds to an $rms$ velocity close to the sound speed,   Figure \ref {fig:compared_graphicsV} indicates that the systems naturally evolve toward a subsonic turbulent regime, with \( v_{\rm rms} / c_s < 1 \), typically ranging from ~0.2 to 0.7 depending on the model.}

\begin{figure}
    \centering
    \includegraphics[width=1.0\columnwidth]{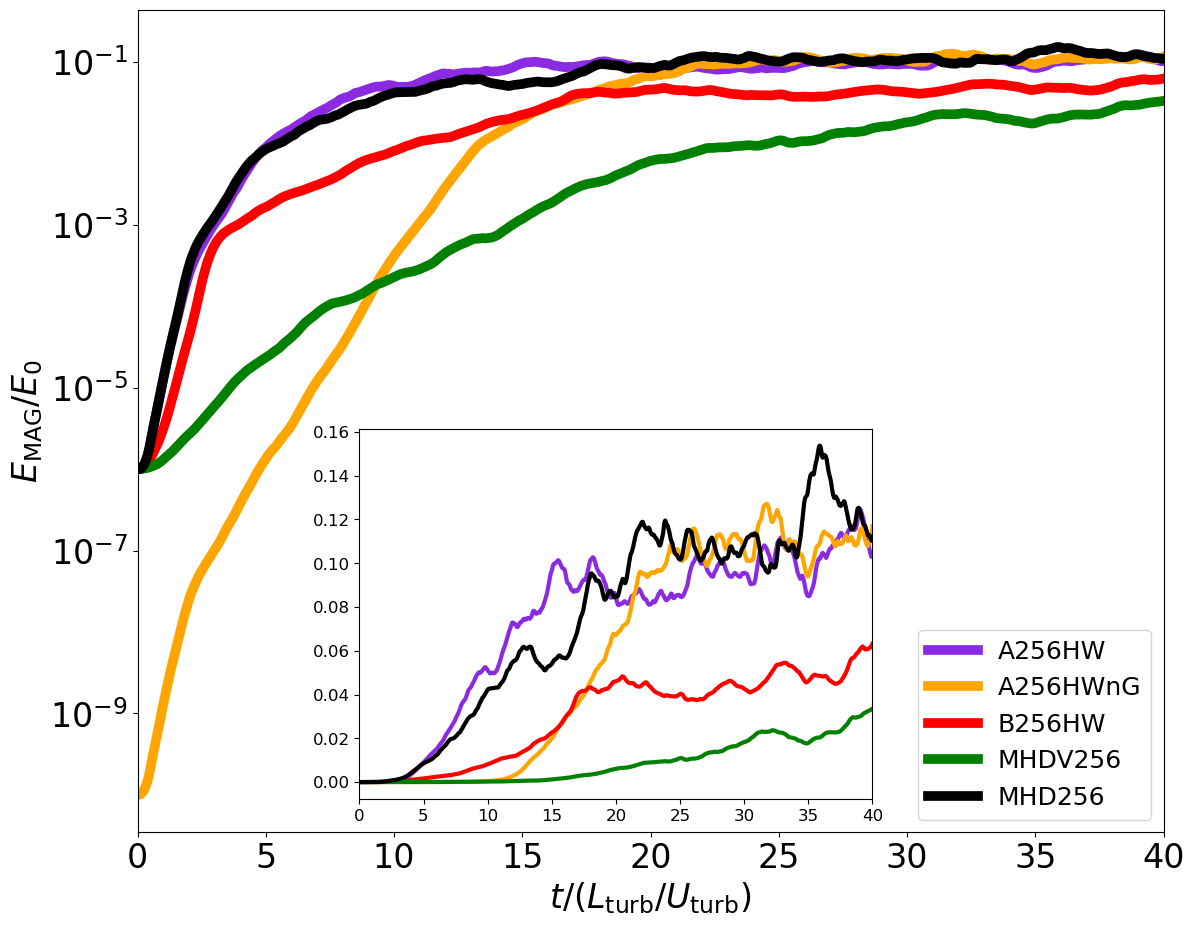}
    \caption{Temporal evolution of the mean magnetic energy density, normalized by the reference kinetic energy density $E_{0} = \frac{1}{2} \rho_{0} U_{\rm turb}^2$, for models A256HW, A256HWnG, B256HW, MHDV256, and MHD256. 
    The main plot presents the energy density on a logarithmic scale, while the inset displays it on a linear scale.
    Model parameters are listed in Table~\ref{tab:initial_condition}.}
    \label{fig:compared_graphics}
\end{figure}

Figure~\ref{fig:compared_graphics} shows the mean magnetic energy density evolution. The run A256HW exhibits an initial exponential growth, reaching approximately $E_M \sim 5 \times 10^{-4} \; E_0$ at $t \sim 3 \; (L_{\rm turb}/U_{\rm turb})$. This is followed by a linear phase until $t \sim 17 \; (L_{\rm turb}/U_{\rm turb})$, when it reaches $E_M \sim 10^{-1} \; E_0$. In the counterpart model with a much smaller initial seed magnetic field,  A256HWnG, the exponential expansion starts later and the linear phase extends until $t \sim 20 \; (L_{\rm turb}/U_{\rm turb})$, also achieving similar $E_M$ values at saturation. The run B256HW follows an evolution similar to that of A256HW, but with saturation at a lower energy level, around $4 \times 10^{-2} \; E_0$. In contrast, run MHDV256 shows much slower growth, reaching $E_M \sim 2 \times 10^{-2} \; E_0$ at the end of the simulation, while run MHD256 reaches values close to A256HW, with slightly delayed saturation.

{\color{black}
Figure \ref{fig:compared_graphics} also indicates that the magnetic energy density saturates at approximately 1–10\% of the initial kinetic energy (\( E_{\rm mag} / E_0 \sim 0.01 - 0.1 \)). Given \( E_0 = \frac{1}{2} \rho_0 c_s^2 \) and \( p_{\rm th} \sim \rho_0 c_s^2 \), we estimate \( \langle \beta \rangle \sim P_{\rm th} / P_{\rm mag} \sim 20 - 200 \). Specifically, $\langle \beta \rangle \simeq$ 20 at magnetic saturation for the ideal MHD and A models, $\langle \beta \rangle \simeq$  200 for the viscous isotropic MHDV model, and has an intermediate value of $\langle \beta \rangle \simeq$ 33 for the B model. While these values for A and B models are somewhat lower than the typical average $\beta \simeq$  100 predicted for the ICM \citep[e.g.,][]{sarazin2003hot}, they align with values found in regions of strong magnetic fields such as cluster cores and filaments \citep[e.g.,][]{vogt2005bayesian}. This behavior is further illustrated in Figure~\ref{fig:beta_evolution} in  Appendix~\ref{sec:appendix_beta}, which shows the time evolution of the volume-averaged plasma $\beta$ for all models. }

\begin{figure*}
    \centering
    \includegraphics[trim=88 58 10 0, clip, width=2.15\columnwidth]{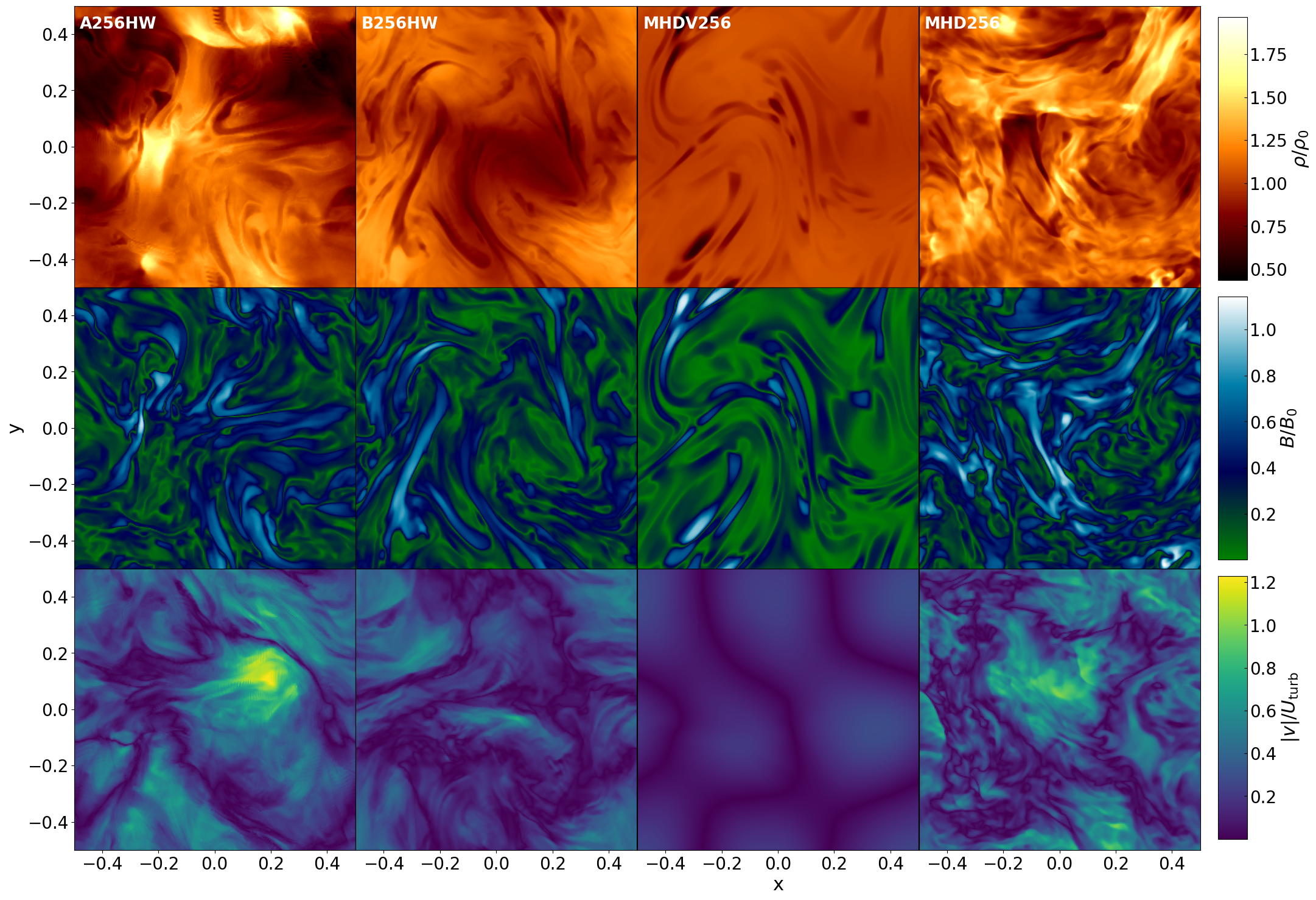}
    \caption{Final distributions of density (top row), magnetic field intensity (middle row), and velocity magnitude (bottom row) in the central slice of models A256HW  (first column), B256HW  (second column),  MHDV256 (third column), and MHD256  (fourth column). The initial seed field $\mathbf{B}_0$ is oriented horizontally.}
    \label{fig:maps_2D}
\end{figure*}

Figure~\ref{fig:maps_2D} exhibits the structures of density, magnetic fields, and velocities in the models during the saturation phase. Among all models, the run A256HW presents a density distribution with well-defined, large-scale structures, a magnetic field spread broadly and in a complex manner throughout the simulated volume, and a dynamic and varied velocity field. 
 The model B256HW exhibits similar structural features, with density, magnetic field, and velocity distributions following the same overall patterns observed in A256HW.

The run MHDV256 presents substantially reduced complexity in density structures, with nearly no spatial variation. 
The magnetic field also shows less complex structures. 
The velocity distribution is almost homogeneous, as isotropic viscosity damps velocity fluctuations, leading to a calmer and less structured system. On the other hand, the run MHD256, which lacks explicit viscosity, exhibits smaller, more numerous density structures, characterized by finer density scales associated with a smaller viscous scale. The magnetic field manifests on smaller scales, while the velocity field shows greater variation at small scales, with smaller structures and irregular flows, characterizing a more chaotic dynamic, as in models A and B.

\subsection{Power Spectra}

\begin{figure}
    \centering
    \includegraphics[width=1.0\columnwidth]{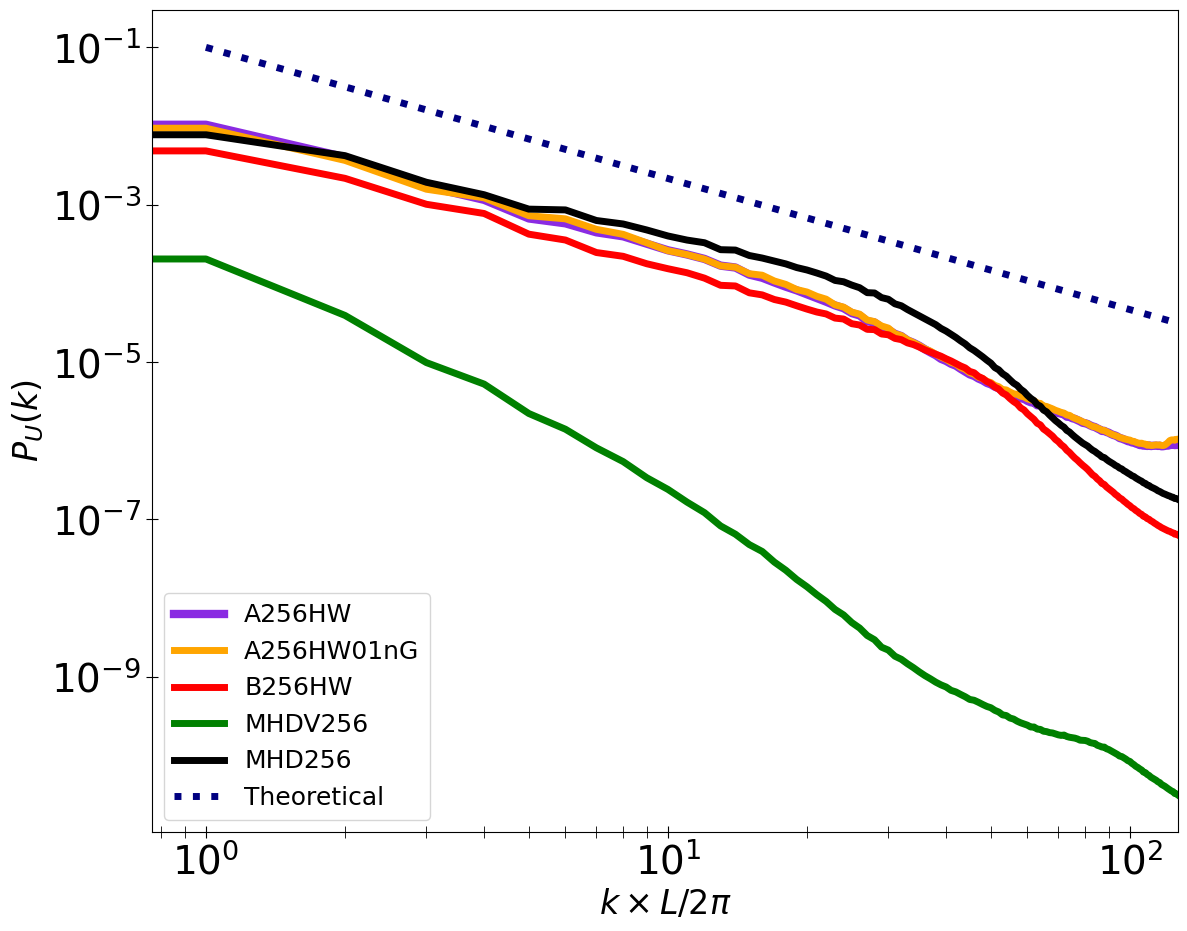} \\
    \includegraphics[width=1.0\columnwidth]{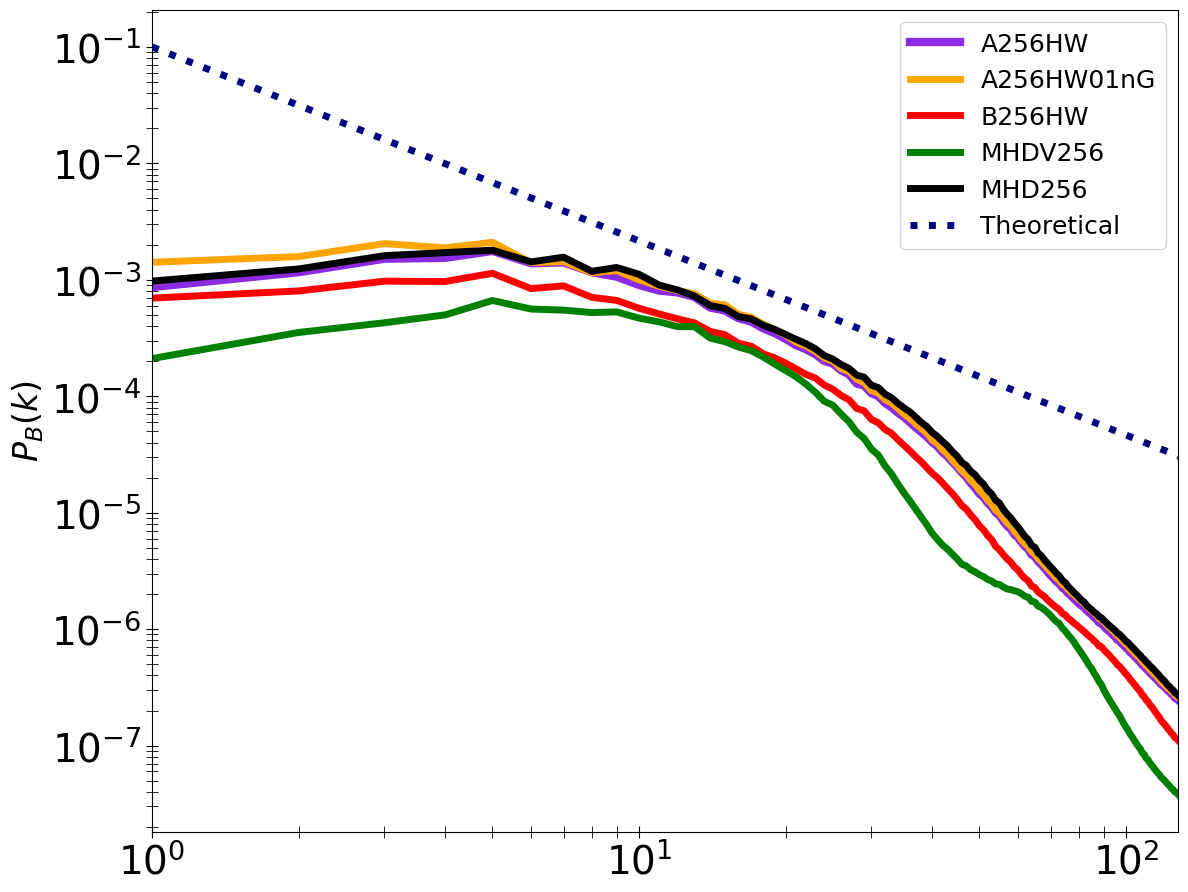}
    \caption{Power spectra of the velocity (top) and magnetic field (bottom) at the final time, $t = 40 L_{\rm turb}/U_{\rm turb}$, for the models A256HW, A256HWnG, B256HW, MHDV256, and MHD256.  Dark blue dotted lines represent theoretical curves for the Kolmogorov spectrum ($\propto k^{-5/3}$).}
    \label{fig:power_spectrum_final}
\end{figure}

Figure~\ref{fig:power_spectrum_final} compares the power spectra of velocity and magnetic fields between the models at their saturated stages ($t = 40 \, L_{\rm turb}/U_{\rm turb}$). The run A256HW shows a velocity spectrum following closely a Kolmogorov spectral index in the inertial range from $k \approx 1 \times 2\pi/L$ to the viscous scale at $k \approx 40 \times 2\pi/L$.
The magnetic field spectrum peaks at $k \approx 5 \times 2\pi/L$. The run A256HWnG is almost indistinguishable from A256HW. The run B256HW has a slightly shallower slope in the velocity spectrum, and the inertial range extends up to $k \approx 50 \times 2\pi/L$.  The magnetic field spectrum also peaks at $k \approx 5 \times 2\pi/L$, but with an amplitude marginally lower than A256HW.

The run MHDV256 does not exhibit an inertial range, with the velocity spectrum amplitude approximately 2 orders of magnitude lower in the largest scales ($k \sim 1  \times 2\pi/L$) than the other models. The shape of the magnetic spectrum is similar to the others, but with a smaller amplitude, between $5 \times 2\pi/L < k < 20 \times 2\pi/L$. 
In contrast, the run MHD256 presents a velocity spectrum with an inertial range comparable to model A256HW, also extending from $k \approx 1 \times 2\pi/L$ to $k \approx 40 \times 2\pi/L$. The magnetic field spectrum is indistinguishable from that of the run A256HW.

\begin{figure*}
    \centering
    \includegraphics[width=2.12\columnwidth]{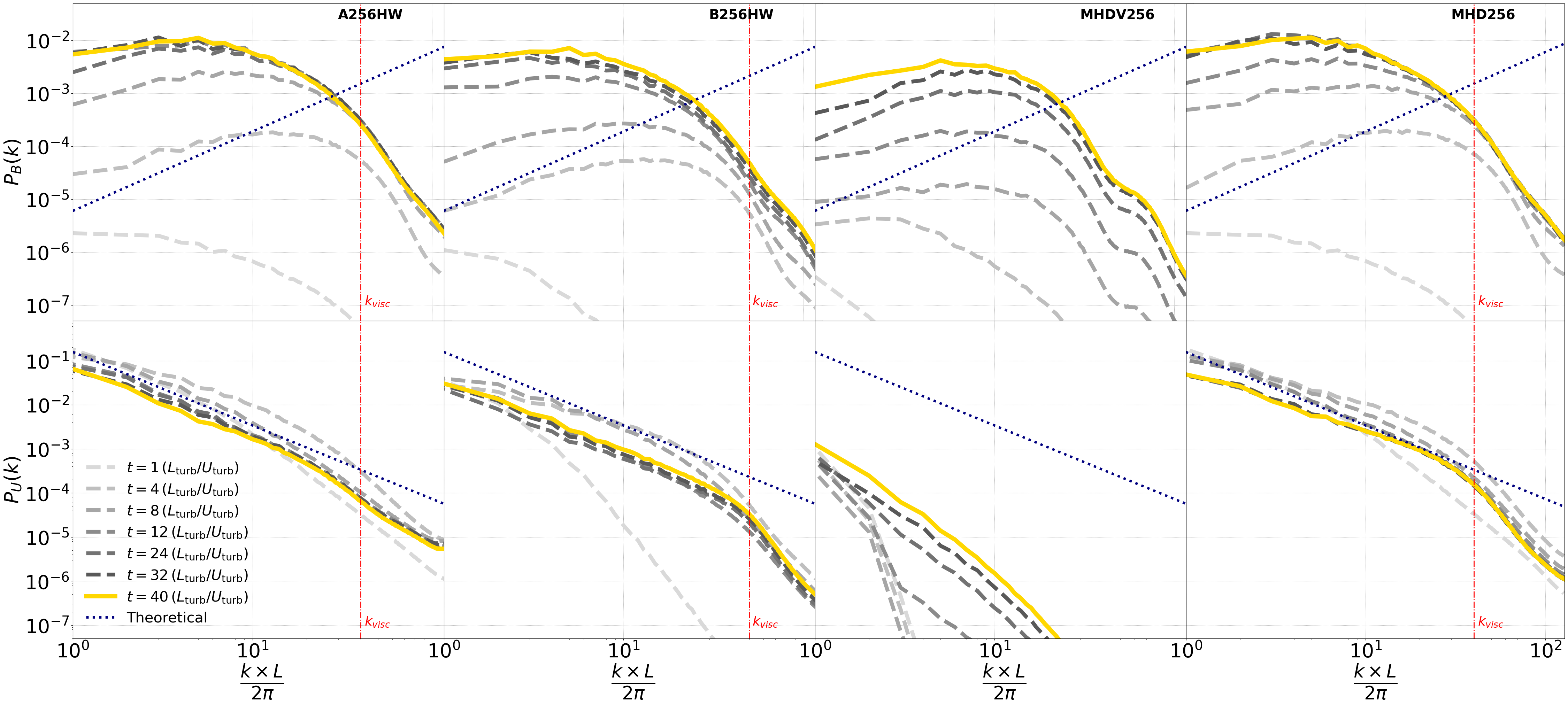}
    \caption{Evolution of the magnetic (top) and the velocity field (bottom) power spectra for the models A256HW (first column), B256HW (second column), MHDV256 (third column), and MHD256 (fourth column),  from $t = 1 \, (L_{\rm turb}/U_{\rm turb})$ to $t = 40 \, (L_{\rm turb}/U_{\rm turb})$. Dark blue dotted lines represent the power laws for the Kolmogorov spectrum ($\propto k^{-5/3}$) in the bottom panels and the Kazantsev spectrum ($\propto k^{3/2}$) in the upper panels. The dashed-dotted red vertical line represents the viscous scale at $t=$ $40 (L_{\rm turb}/ U_{\rm turb})$ for each model.}
    \label{fig:power_spectrum}
\end{figure*}

Figure~\ref{fig:power_spectrum} shows the temporal evolution of the power spectra of the magnetic field (upper panels) and the velocity field (lower panels). 
In the run A256HW, the magnetic field spectrum at larger scales ($k < 3 \times 2\pi/L$) is roughly consistent with the Kazantsev spectrum (shown in dotted line) in the initial stages ($t \sim 4 L_{\rm turb}/U_{\rm turb}$). From $t = 8 L_{\rm turb}/U_{\rm turb}$, the power spectrum saturates at smaller scales ($k = 20 \times 2\pi/L$), initiating field growth at larger scales (linear dynamo phase) until $t \sim 24 L_{\rm turb}/U_{\rm turb}$. At $t \sim 4 L_{\rm turb}/U_{\rm turb}$, the velocity spectrum exhibits a distinct curvature in the range $1 \times 2\pi/L \lesssim k \lesssim 10 \times 2\pi/L$, characterized by a slight concavity.
In run B256HW, the magnetic field spectrum initially is close to the Kazantsev spectrum up to $k \approx 3 \times 2\pi/L$, reaching the linear dynamo phase at $t \sim 12 L_{\rm turb}/U_{\rm turb}$ and saturation after $t \sim 24 L_{\rm turb}/U_{\rm turb}$. The velocity spectrum exhibits a curvature profile at $t \sim 4 L_{\rm turb}/U_{\rm turb} - 8 L_{\rm turb}/U_{\rm turb}$ similar to the run A256HW but with a more pronounced concavity, developing an inertial range consistent with the Kolmogorov scaling between $ 2 \times 2\pi/L \lesssim k < 40 \times 2\pi/L$ already by $t \sim 12 L_{\rm turb}/U_{\rm turb}$. 

The run MHDV256 exhibits slower initial amplification of the magnetic field on larger scales, and the profile in the initial states shows no resemblance to the Kazantsev spectrum. The velocity spectrum gradually increases due to forced turbulence but does not develop an inertial range. The run MHD256, on the other hand, reaches the linear dynamo phase at $t \sim 8 L_{\rm turb}/U_{\rm turb}$ and the saturated phase at $t \sim 24 L_{\rm turb}/U_{\rm turb}$ similar to runs A256HW and B256HW. The velocity spectrum exhibits a slope change only at $t \sim 16 L_{\rm turb}/U_{\rm turb}$, with an inertial range similar to that of the final spectrum.

\subsection{Evolution of the Viscosity}

\begin{figure*}
    \centering
    \includegraphics[trim=260 138 200 130, clip, width=0.640\columnwidth]{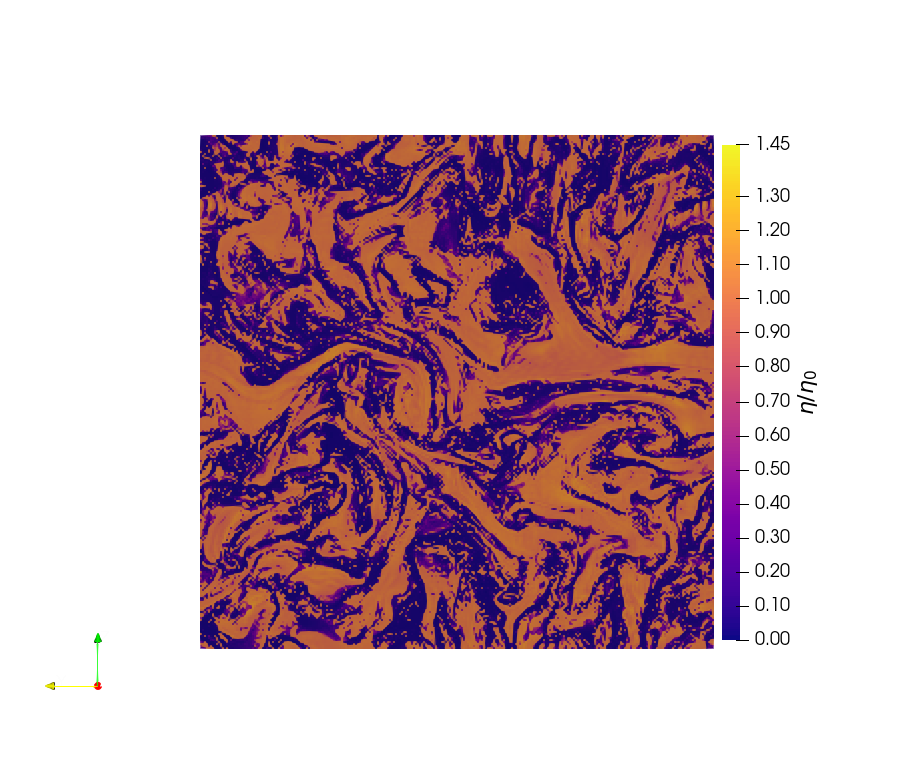}
    \includegraphics[trim=260 138 200 130, clip, width=0.640\columnwidth]{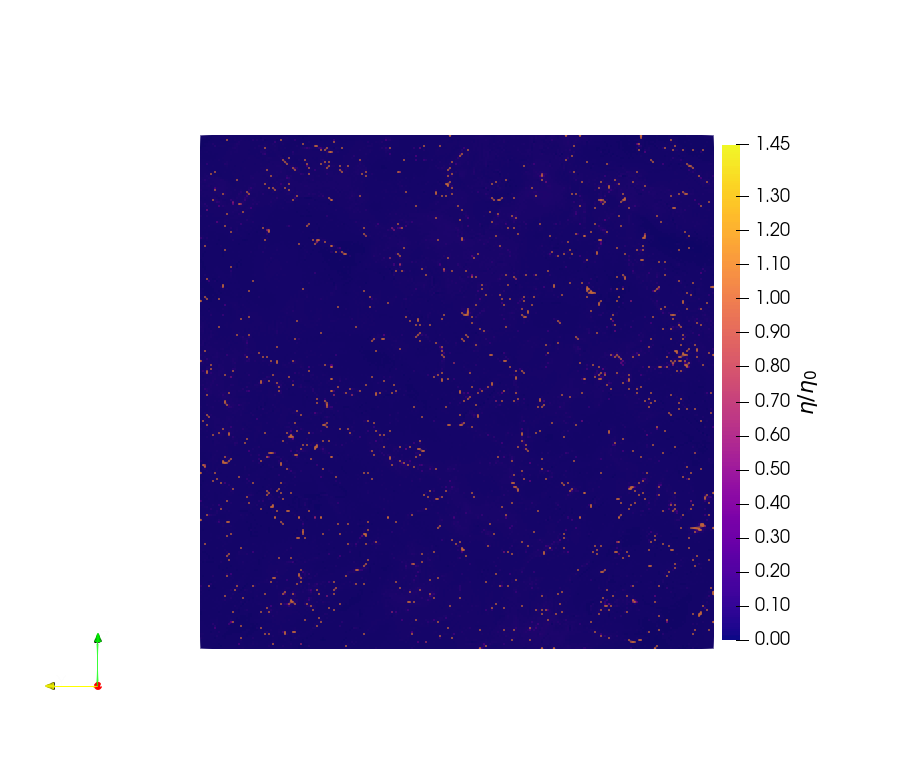}
    \includegraphics[trim=260 138 90 130, clip, width=0.795\columnwidth]{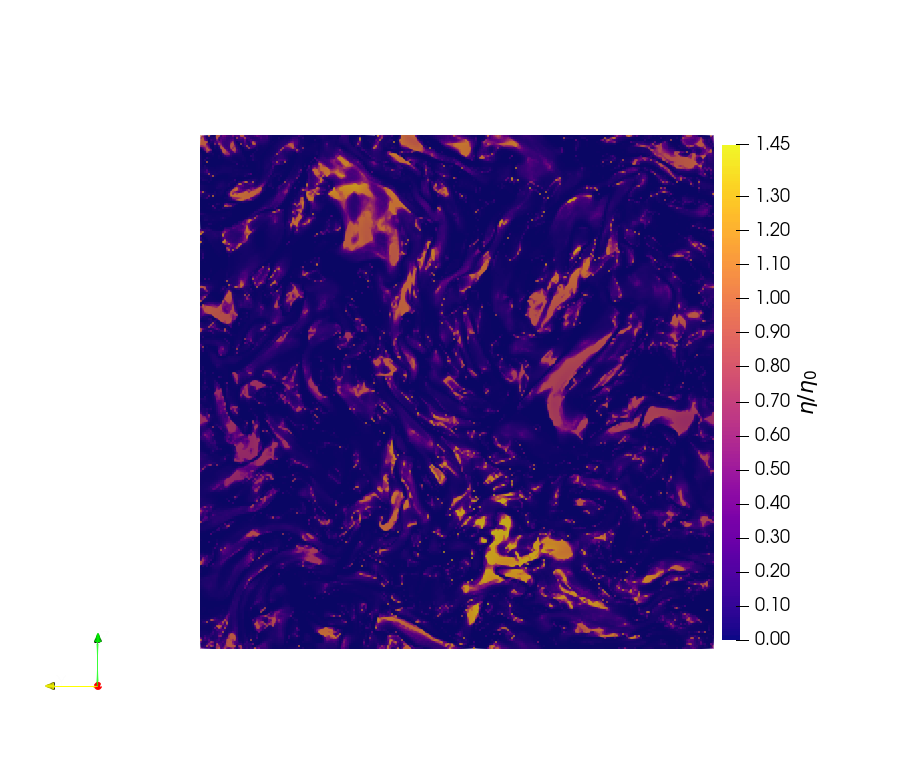}
    \includegraphics[trim=260 138 200 130, clip, width=0.640\columnwidth]{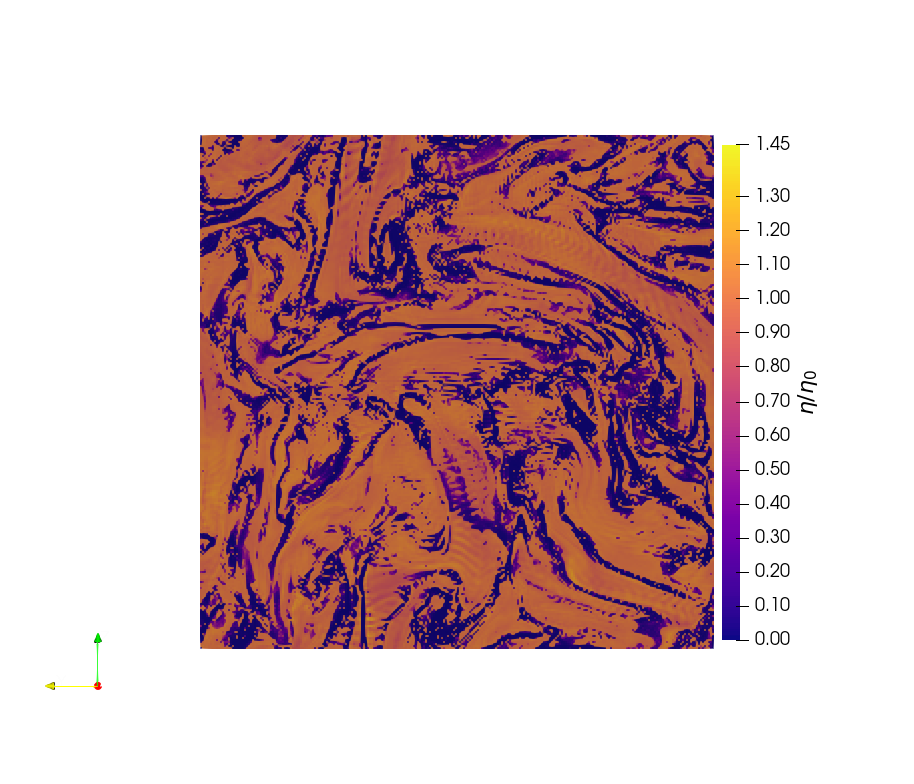}
    \includegraphics[trim=260 138 200 130, clip, width=0.640\columnwidth]{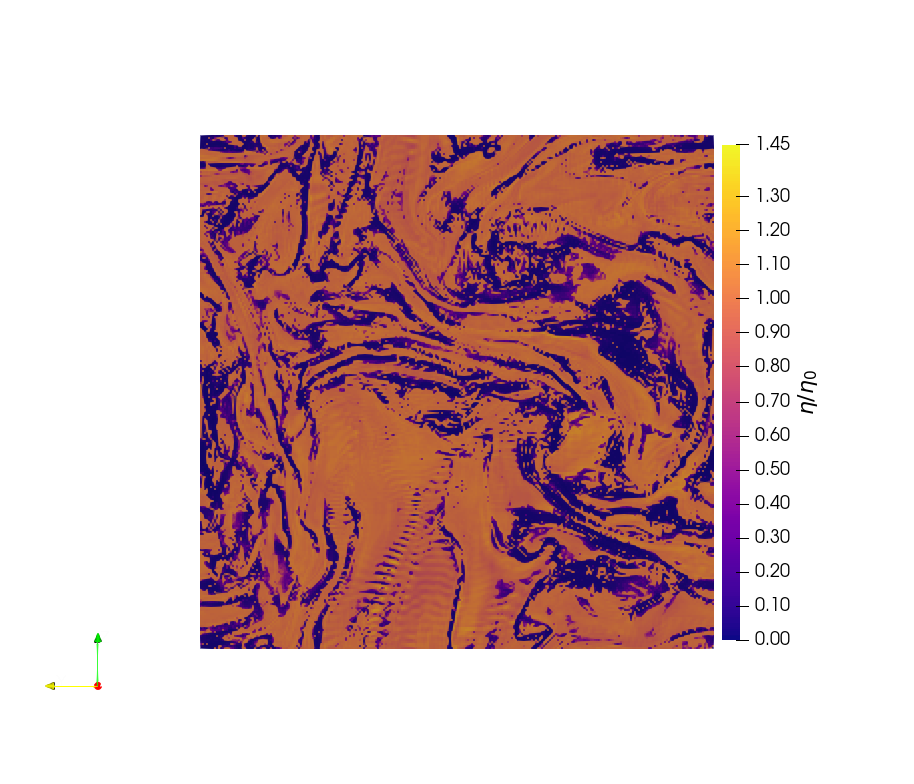}
    \includegraphics[trim=260 138 90 130, clip, width=0.795\columnwidth]{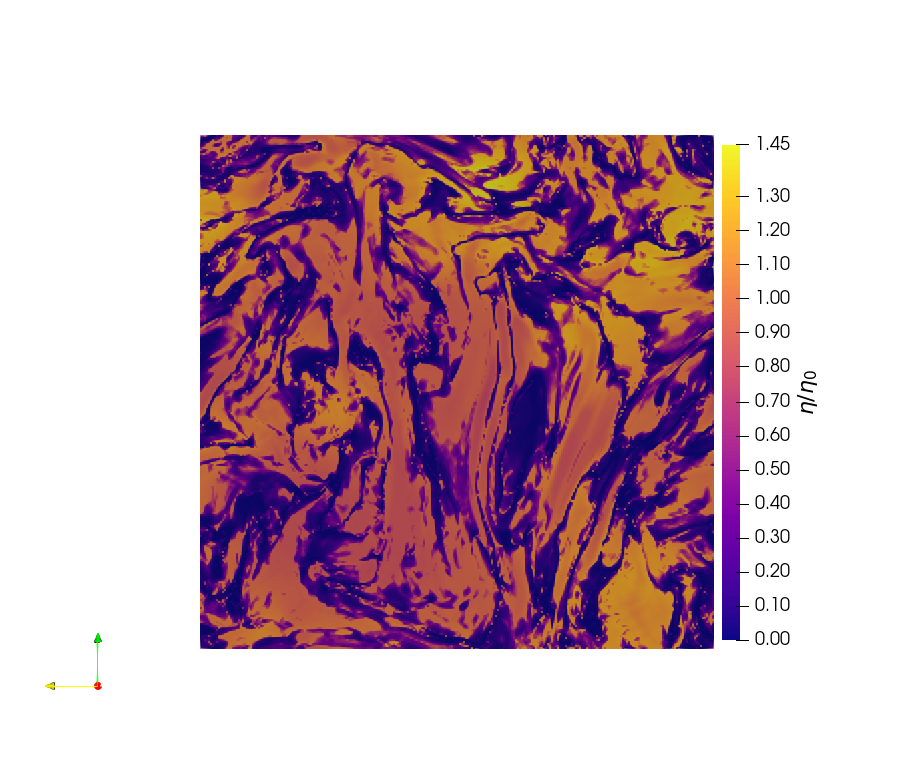}
    \caption{Central slice showing the distribution of the effective kinematic viscosity ($\eta/\eta_0$) for the A256HW (left column), A256HWnG (middle column), B256HW (right column) models. The maps correspond to two-time snapshots: $t = 10 L_{\rm turb}/U_{\rm turb}$ (top) and $t = 40 L_{\rm turb}/U_{\rm turb}$ (bottom). The initial seed field \textbf{B$_0$} is oriented horizontally.}
\label{fig:mapa_coeficiente_viscoso}
\end{figure*}

The bottom panels of Figure \ref{fig:mapa_coeficiente_viscoso} present the distributions of the effective kinematic viscosity ($\eta/\eta_0$) at the final time for different runs. Both A models exhibit a highly non-uniform distribution of $\eta/\eta_0$, with lower viscosity regions highlighted by dark elongated filamentary structures, and high viscosity areas represented by light, wider ribbons.  In contrast, low-viscosity regions are more prominent in the run B256HW (bottom panels), displaying broader dark filaments.

The top panels of Figure \ref{fig:mapa_coeficiente_viscoso} show the same $\eta/\eta_0$ maps as the bottom panels but at an earlier time $t = 10 L_{\rm turb}/U_{\rm turb}$. The differences between the runs A256HW, A256HWnG, and B256HW, reflect a different temporal evolution of this distribution.  As illustrated in Figure \ref{fig:compared_graphics}, at this time, the model A256HW has reached an average amplitude of $\left( \frac{B^2/8\pi}{E_0} \right) \sim 0.06$, while model A256HWnG is still at $\left( \frac{B^2/8\pi}{E_0} \right) \sim 0$ and the model B256HW has reached $\left( \frac{B^2/8\pi}{E_0} \right) \sim 0.01$.

\begin{figure*}
    \centering
    \includegraphics[width=2.0\columnwidth]{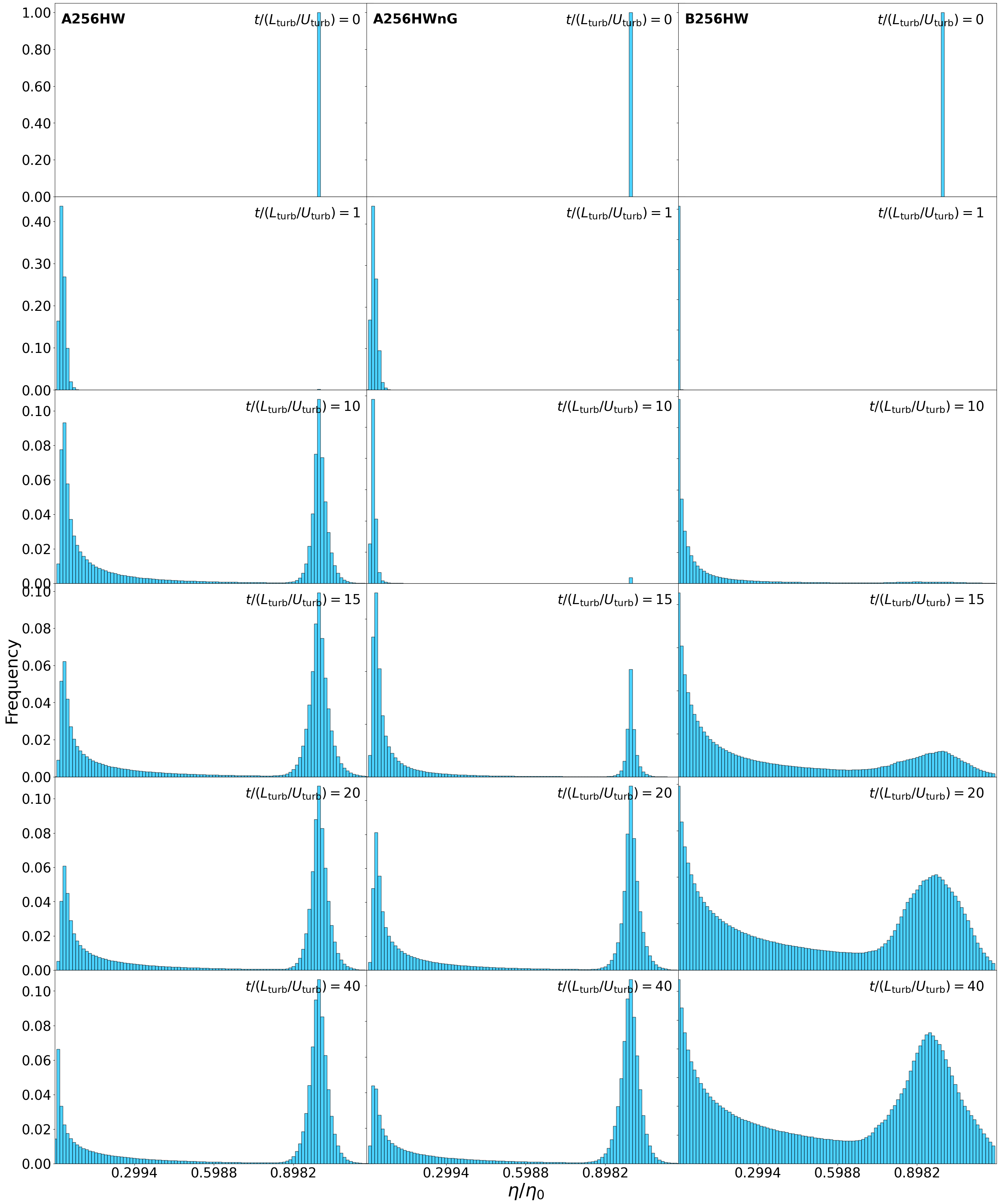}
    \caption{Distribution of the effective viscosity coefficients $\eta/\eta_0$,
    shown at different times for models A256HW (left column), A256HWnG (middle column), and B256HW (right column). Each row represents a different times: $t = 0$, 1, 10, 15, 20, and 40 $(L_{\rm turb}/U_{\rm turb})$. We call attention to the different vertical scales in different times.}
    \label{fig:evolution_histogram_coefVisc}
\end{figure*}

Figure \ref{fig:evolution_histogram_coefVisc} presents histograms showing the evolution of the distribution of the effective viscosity coefficient $\eta/\eta_0$. For the run A256HW (left column), at $t = 0  (L_{\rm turb}/U_{\rm turb})$, the distribution reflects the initial value imposed on the system. After one turbulence turnover time ($t = 1  (L_{\rm turb}/U_{\rm turb})$), there is a reduction of the coefficient that peaks around $10^{-3}$. As time progresses ($t = 10, 15, 20$, and $40  (L_{\rm turb}/U_{\rm turb})$), the distribution spreads, and forms a bimodal profile. The evolution is similar for run A256HWnG (middle column), but the bimodal profile appears later, only from $t \sim 20  (L_{\rm turb}/U_{\rm turb})$ onward when the magnetic field amplitudes in the two models also become similar as both have reached saturation (see Fig. \ref{fig:compared_graphics}). The run B256HW (right column) shows more distinct features.  The bimodal distribution which appears around $t \sim 15  (L_{\rm turb}/U_{\rm turb})$ is broader and smoother than in runs A256HW and A256HWnG, reflecting a smoother evolution of the viscosity properties.

The appearance of the bimodal distribution in the effective viscosity coefficient ($\eta$) arises from the interaction between forced turbulence and the system's tendency to return to equilibrium. Initially, $\eta$ is uniform and consistent with a physical equilibrium. The injection of forced turbulence rapidly decreases $\eta$, leading to marginally unstable regions characterized by low viscosity. As turbulence evolves within the system, a fraction of the plasma returns to the initial value, while another remains marginally unstable due to continuous energy injection, sustaining the conditions necessary for firehose and mirror instabilities. This results in bimodality, with one peak associated with the stable state and the other with turbulent regions.

\begin{figure}
    \centering
    \includegraphics[width=1.0\columnwidth]{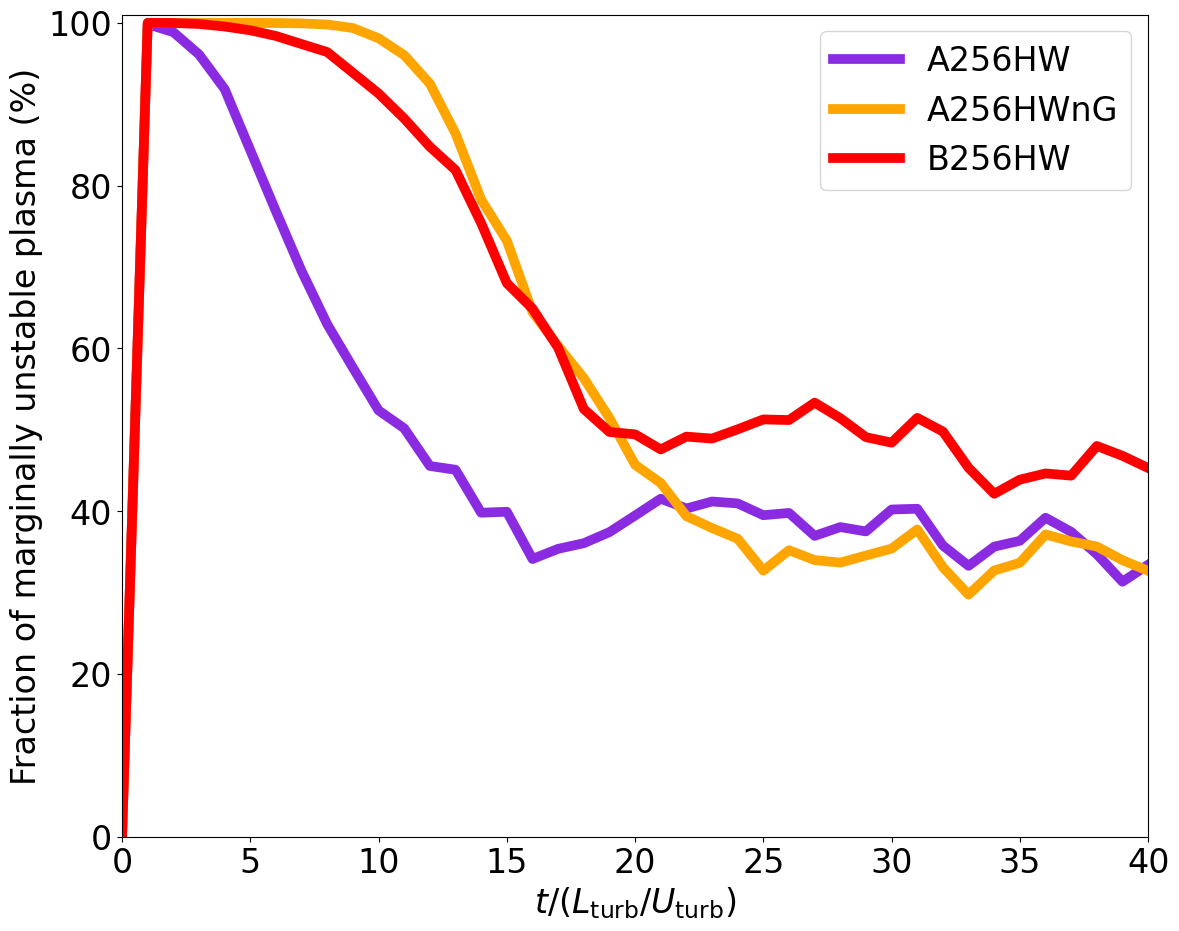}
    \caption{Evolution of the fraction of plasma at the instabilities threshold for models A256HW (purple curve), A256HWnG (orange curve), and B256HW (red curve). It highlights the marginally unstable plasma subset.}
    \label{fig:evolution_moda_coefVisc_below01_percent}
\end{figure}

\begin{figure*}
    \centering
    \includegraphics[width=2.0\columnwidth]{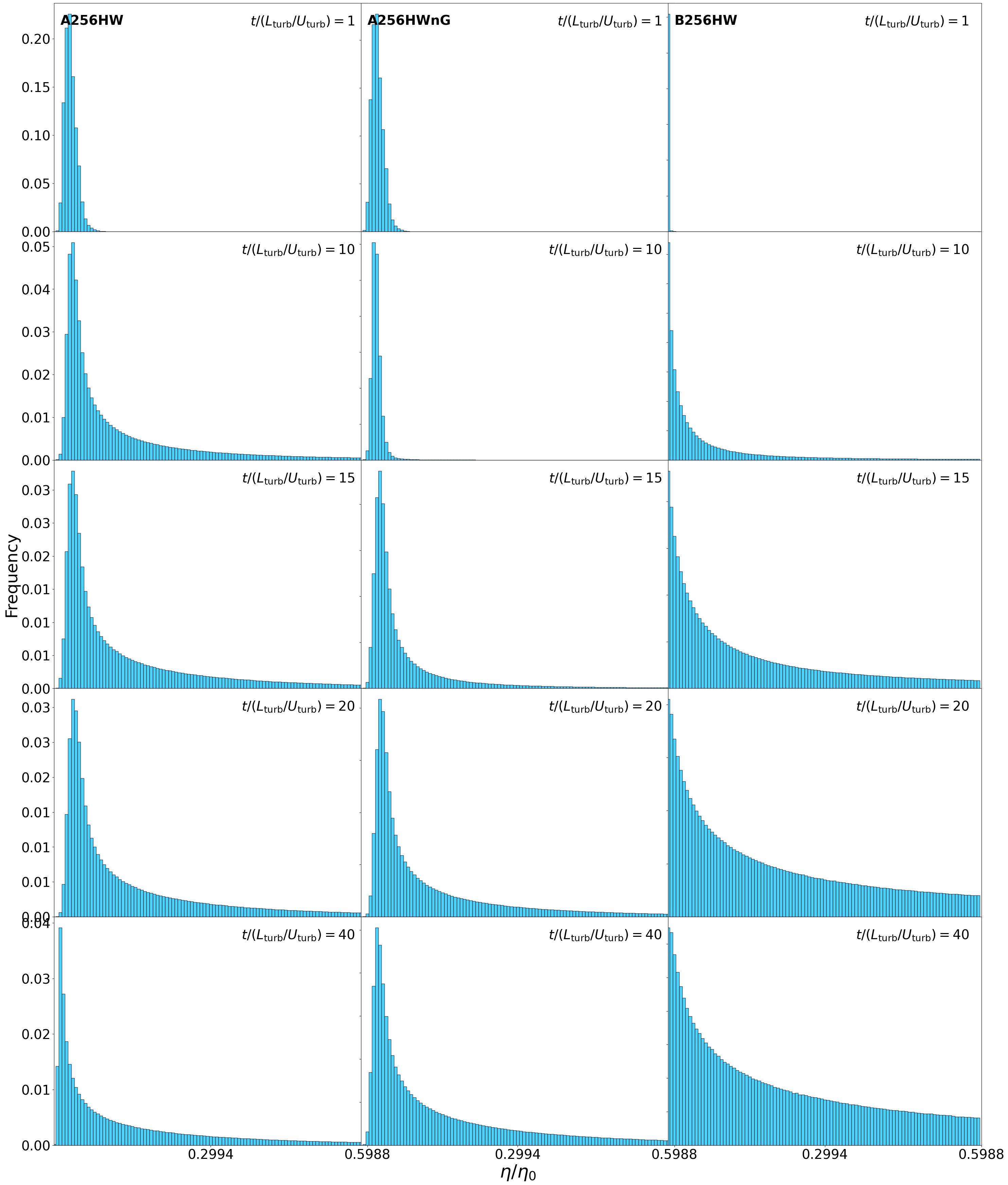}
    \caption{Distribution of the effective viscosity coefficients ($\eta/\eta_0$),    shown in different times for models A256HW (left column), A256HWnG (middle column), and B256HW (right column) for the fraction of marginally unstable plasma. Each row represents a snapshot at different times: $t = 1$, 10, 15, 20, and 40 $(L_{\rm turb}/U_{\rm turb})$.  Attention to different vertical scales at different times.}
    \label{fig:evolution_histogram_coefVisc-instavel}
\end{figure*}

Figures  \ref{fig:evolution_moda_coefVisc_below01_percent} and  \ref{fig:evolution_histogram_coefVisc-instavel} help us to understand the behavior revealed by Figure \ref{fig:evolution_histogram_coefVisc}. 
Figure \ref{fig:evolution_moda_coefVisc_below01_percent} shows the time evolution of the number density fraction of plasma near the firehose and mirror instabilities threshold. This figure highlights the subset of plasma with {\color{black}viscosity}
values below the threshold defined at the saturated stage (Figure \ref{fig:evolution_histogram_coefVisc}). {\color{black}Specifically, we define plasma as unstable when the normalized effective viscosity satisfies $\eta/\eta_0 \lesssim 0.6$, where $\eta_0 = 0.0167$ is the initial kinematic viscosity. This threshold corresponds to the local minimum between the two peaks in the viscosity distribution (see Figure~\ref{fig:evolution_histogram_coefVisc}), separating stable plasma from regions affected by the instabilities where the hardwall limiter is applied. We verified that small changes in this cutoff do not significantly affect the results  due to  the distinct separation between the regimes.}
This density fraction is normalized by the total number density. 

In the early stages ($t < 5  (L_{\rm turb}/U_{\rm turb})$), all models reach marginally unstable plasma fractions of $100\%$, due to the initial injection of turbulence into the system. From $t \sim 10  (L_{\rm turb}/U_{\rm turb})$, the run A256HW exhibits a sharper decline, stabilizing at values close to $40\%$. The run A256HWnG, although similar, stabilizes around $40\%$ starting at $t \sim 25  (L_{\rm turb}/U_{\rm turb})$. In contrast, the run B256HW maintains a consistently higher fraction over time, stabilizing around $50\%$ starting at $t \sim 20  (L_{\rm turb}/U_{\rm turb})$.

{\color{black}\textbf{}}

Figure \ref{fig:evolution_histogram_coefVisc-instavel} shows the distribution of $\eta/\eta_0$ in this marginally unstable plasma subset. 
This figure emphasizes the temporal evolution of the effective viscosity distribution within this fraction. Clearly, it shows a single peak.

\begin{figure}
    \centering
    \includegraphics[width=1.0\columnwidth]{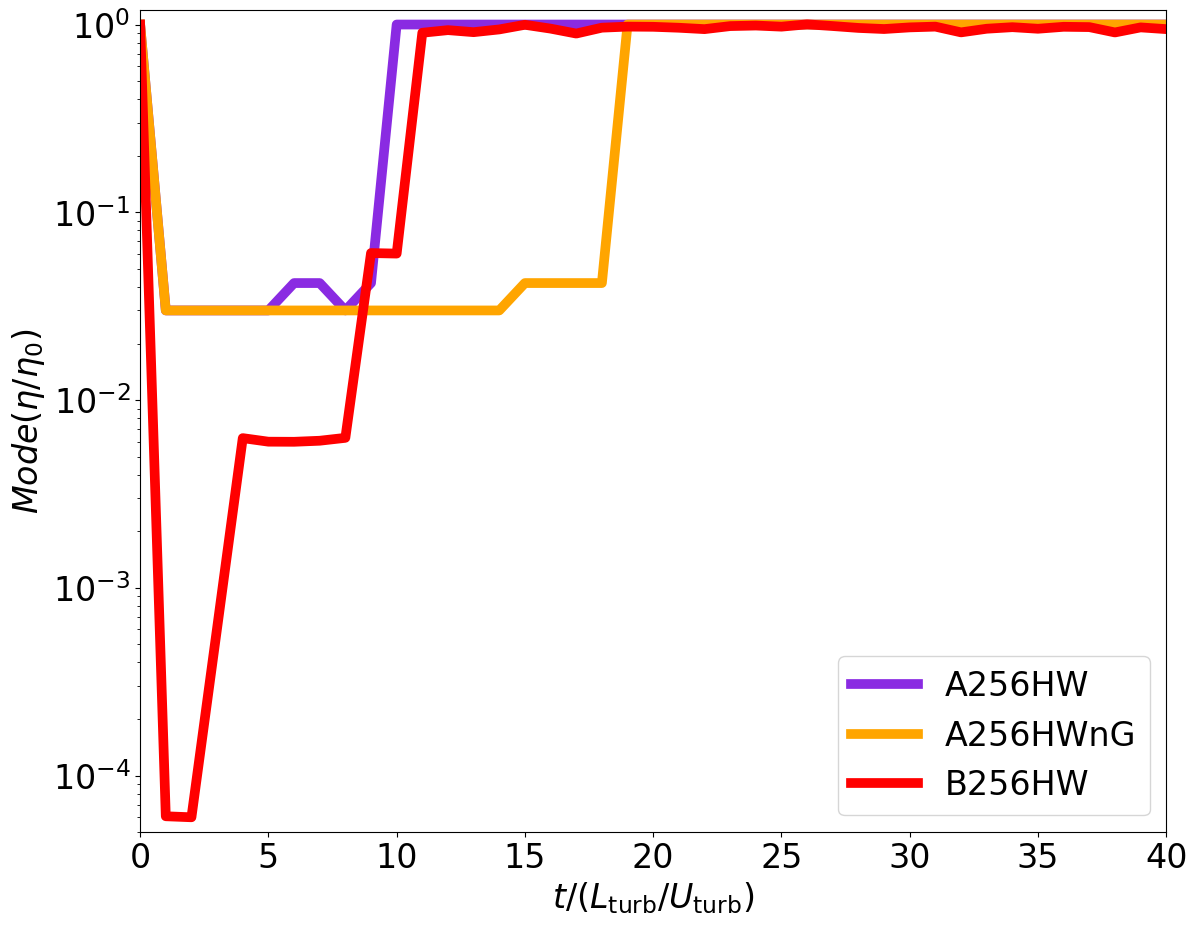}
    \includegraphics[width=1.0\columnwidth]{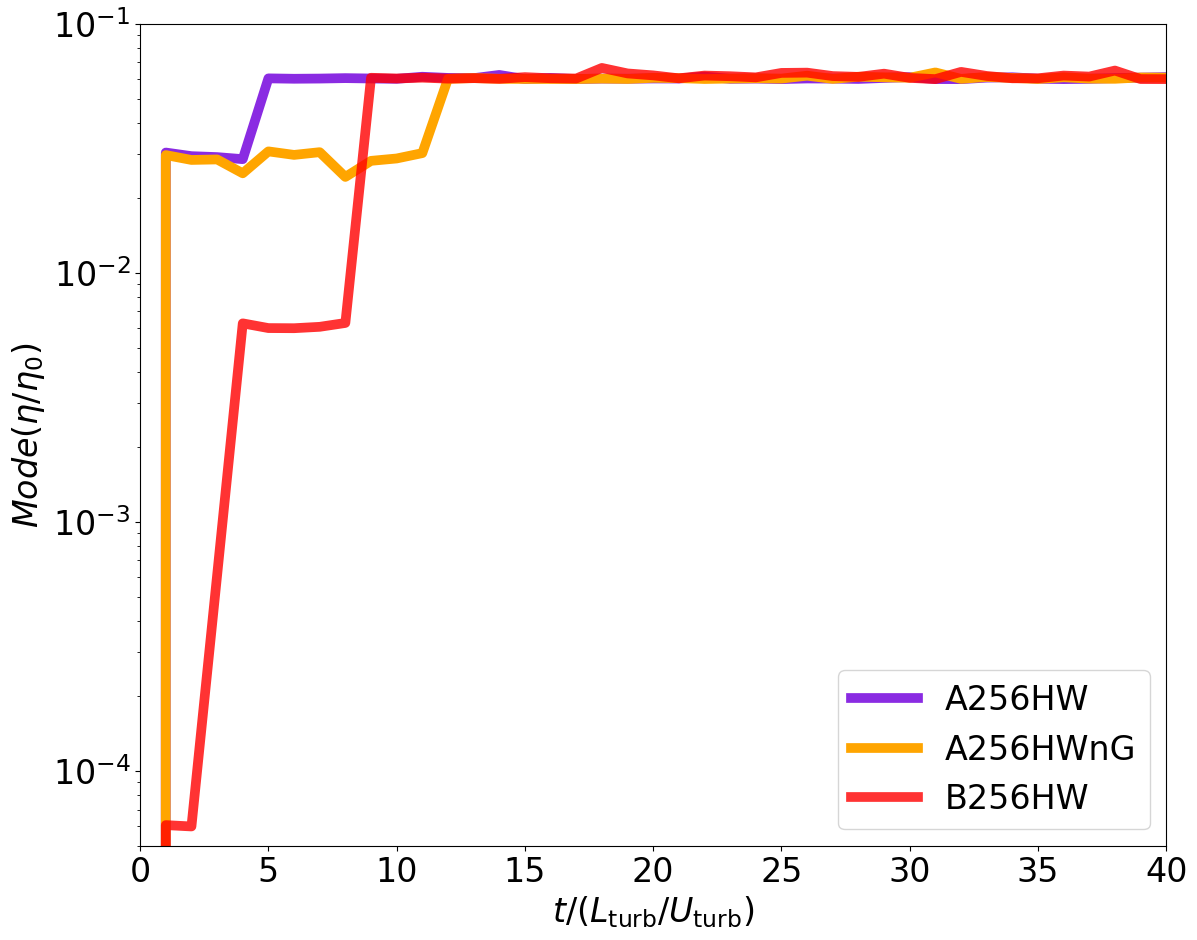}
    \caption{Evolution of the mode of the effective viscosity coefficients for models A256HW (purple curve), A256HWnG (orange curve), and B256HW (red curve). The upper panel shows the evolution of the entire plasma volume, while the lower panel displays the evolution, considering only the fraction of marginally unstable plasma. (Note that the y-axis scale of both diagrams is different.)}
    \label{fig:combined_moda_viscosity}
\end{figure}

Figure \ref{fig:combined_moda_viscosity} shows the time evolution of the mode distribution of the effective viscosity for both the entire system  (top panel) and the marginally unstable fraction of the plasma (bottom panel). This reflects the dominant values in the distribution. In models A, the mode of $\eta$ rapidly drops to $\approx 3 \times 10^{-2} \eta_0$, driven by the marginally unstable plasma fraction shortly after the simulation starts ($t \sim 1 (L_{\rm turb}/U_{\rm turb})$). In the run A256HW (purple line), it starts to increase by $t \sim 10 (L_{\rm turb}/U_{\rm turb})$ and stabilizes near its initial level $\eta_0$. In contrast, the run A256HWnG (orange line) shows a slower evolution, reaching the plateau later.
The run B256HW (red line) exhibits a sharper initial drop in the whole plasma mode distribution, reaching lower values ($\sim 5 \times 10^{-5}$), with unstable regions showing lower average viscosities (see Fig. \ref{fig:evolution_histogram_coefVisc-instavel}). This model also stabilizes, eventually returning to a plateau close to the initial viscosity, at nearly the same time as model A256HW.

The initial sharp drop of the viscosity followed by a period with a minimum value and then, a rapid growth back to the initial (maximum) value, which is seen in both A and B models is consistent with the different phases of the turbulent dynamo action seen in Figure \ref{fig:compared_graphics}.  In the early stage, the continuous injection of forced turbulence into the system triggers kinetic instabilities, significantly reducing viscosity. This reduction coincides with the exponential growth phase of the turbulent dynamo, where strong velocity gradients and fluctuating plasma motions drive rapid magnetic field amplification.

As the dynamo transitions into the linear growth phase, viscosity shows a modest recovery, reflecting a transient balance between the driving instabilities and dissipative effects, coupled with magnetic back-reaction. The gradual rise in viscosity during this phase highlights the interplay between magnetic amplification and dissipation, shaping the transport properties of the plasma.
Finally, the partial recovery of viscosity over time is driven by the damping of instabilities, despite ongoing turbulence. 

For comparison with Figure \ref{fig:combined_moda_viscosity} (top), Figure \ref{fig:coefficient_viscosity_evolucao_log_nub} of Appendix \ref{apdx:viscosity_evolution} presents the time evolution of the $average$ of the effective viscosity coefficients for the entire system. Both diagrams are very similar, as expected.

\begin{figure*}
\centering
\includegraphics[width=1.5\columnwidth]{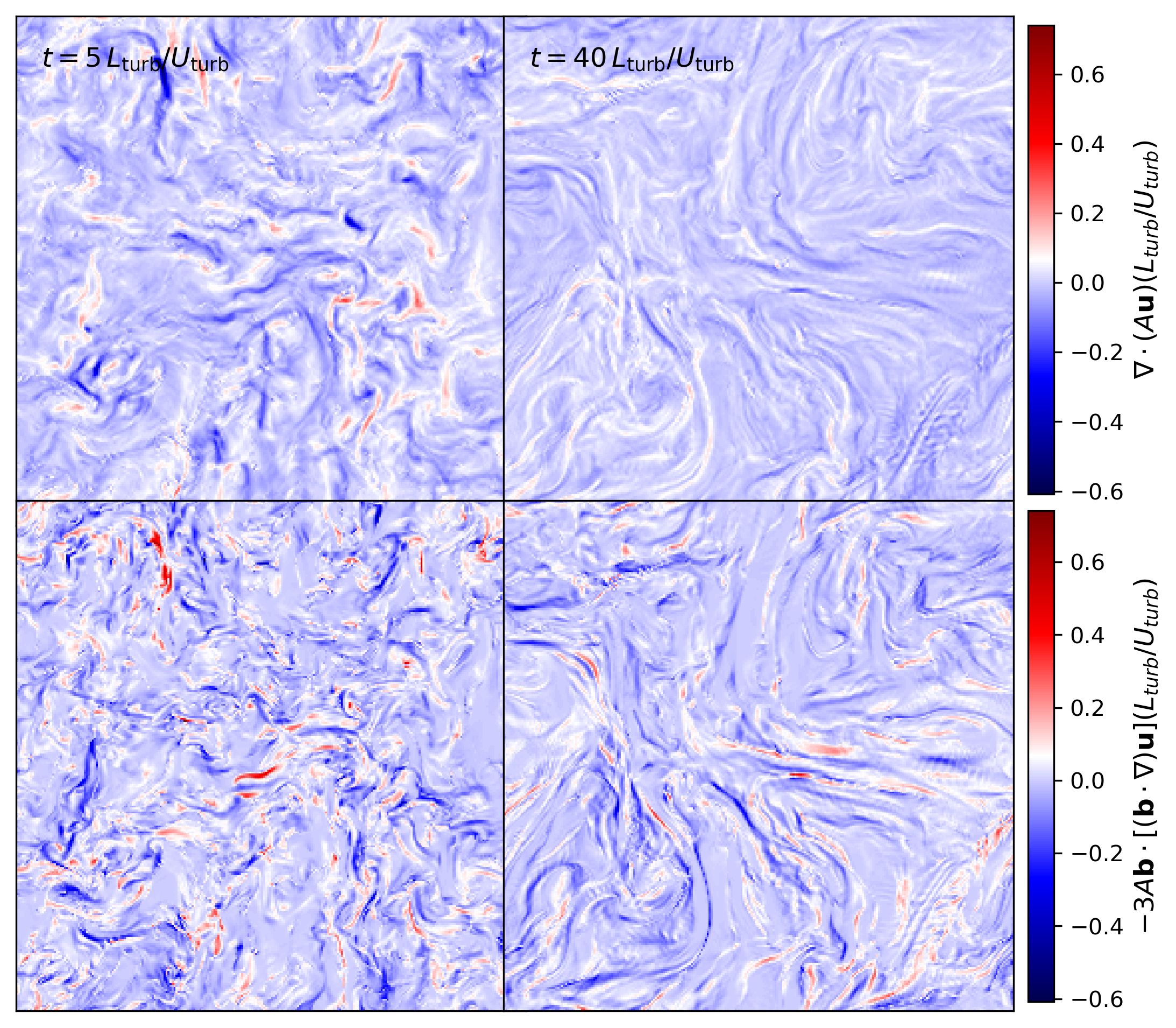}
\caption{Central slice comparison of the divergence term  (top row) and the shear term  (bottom row) from Eq. \ref{eq:dadt_scatt_instab} for model A256HW at $t = 5 \, L_{\rm turb} / U_{\rm turb}$ (left) and $t = 40 \, L_{\rm turb} / U_{\rm turb}$ (right). The initial seed field \textbf{B$_0$} it oriented horizontally.} 
\label{fig:divergence_compression_comparison}
\end{figure*}

Finally, Figure \ref{fig:divergence_compression_comparison} compares the divergence and shear terms from Eq. \ref{eq:dadt_scatt_instab} for the anisotropy time-derivative for model A256HW at $t = 5 \, L_{\rm turb} / U_{\rm turb}$ and $t = 40 \, L_{\rm turb} / U_{\rm turb}$,  to represent an early phase, when turbulence is still developing, and a later stage, dominated by nonlinear effects. These terms are directly linked to the evolution of the effective viscosity. Their comparison quantifies the relative roles of compressive (represented by the divergence term) and shear motions in the turbulence-driven magnetic field amplification. We clearly see that the shear term remains dominant over the divergence term at both times, with this dominance becoming more pronounced at $t = 40 \, L_{\rm turb} / U_{\rm turb}$, where the difference in magnitude between the two terms becomes more evident. 

\subsection{Dependence on Numerical Resolution}

 Numerical resolution plays a fundamental role in accurately capturing viscous structures and magnetic field dissipation in plasma simulations. Insufficient resolution can lead to an underestimation of gradients and suppression of relevant physical processes, affecting the overall dynamics of the system. To assess the robustness of our results, we performed a convergence test, analyzing the evolution of magnetic energy and viscosity structures at different resolutions (\(64^3\), \(128^3\), and \(256^3\)). This analysis allows us to verify the stability of the physical diagnostics and justify the adoption of the \(256^3\) resolution in the main simulations.  

\begin{figure}[!htb]
    \centering
    \includegraphics[width=1.0\columnwidth]{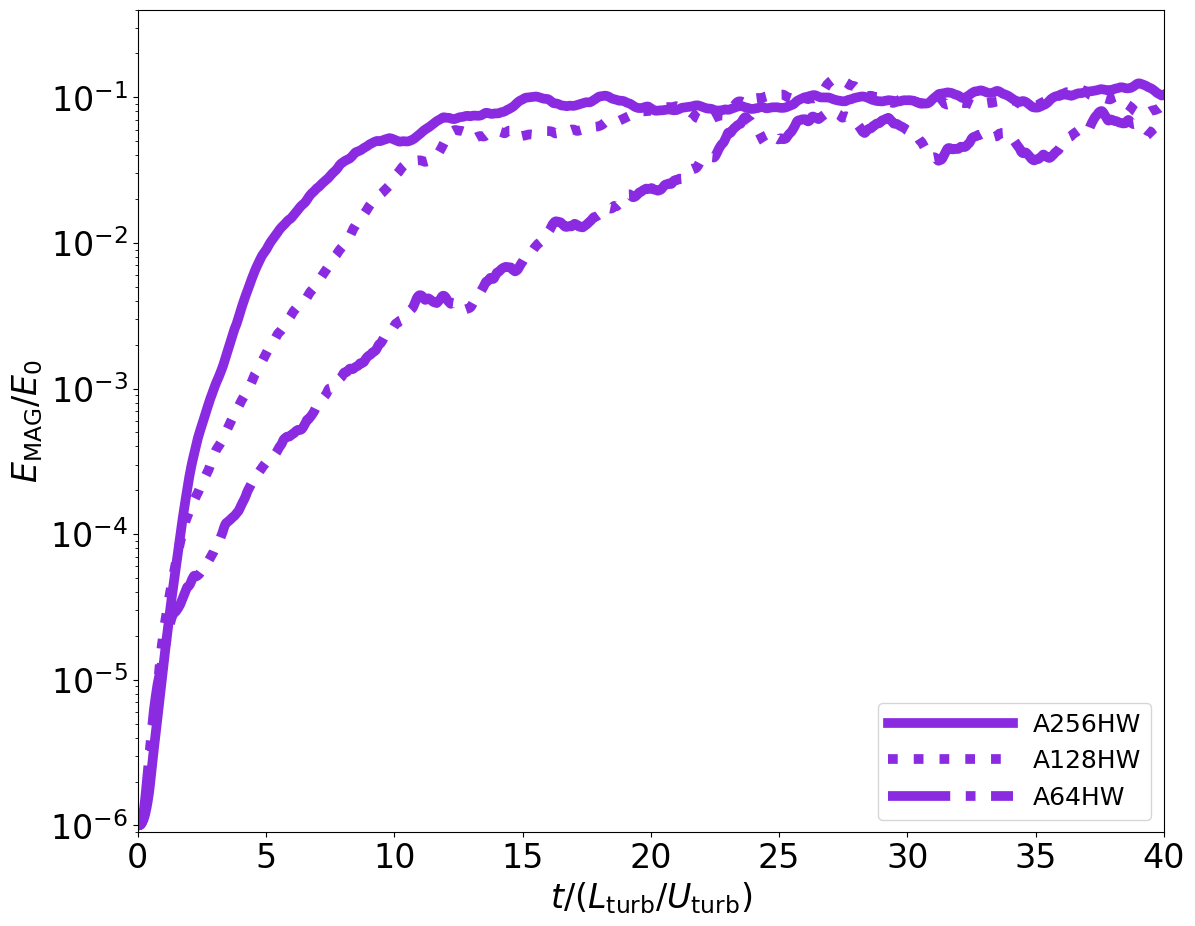}
    \includegraphics[width=1.0\columnwidth]{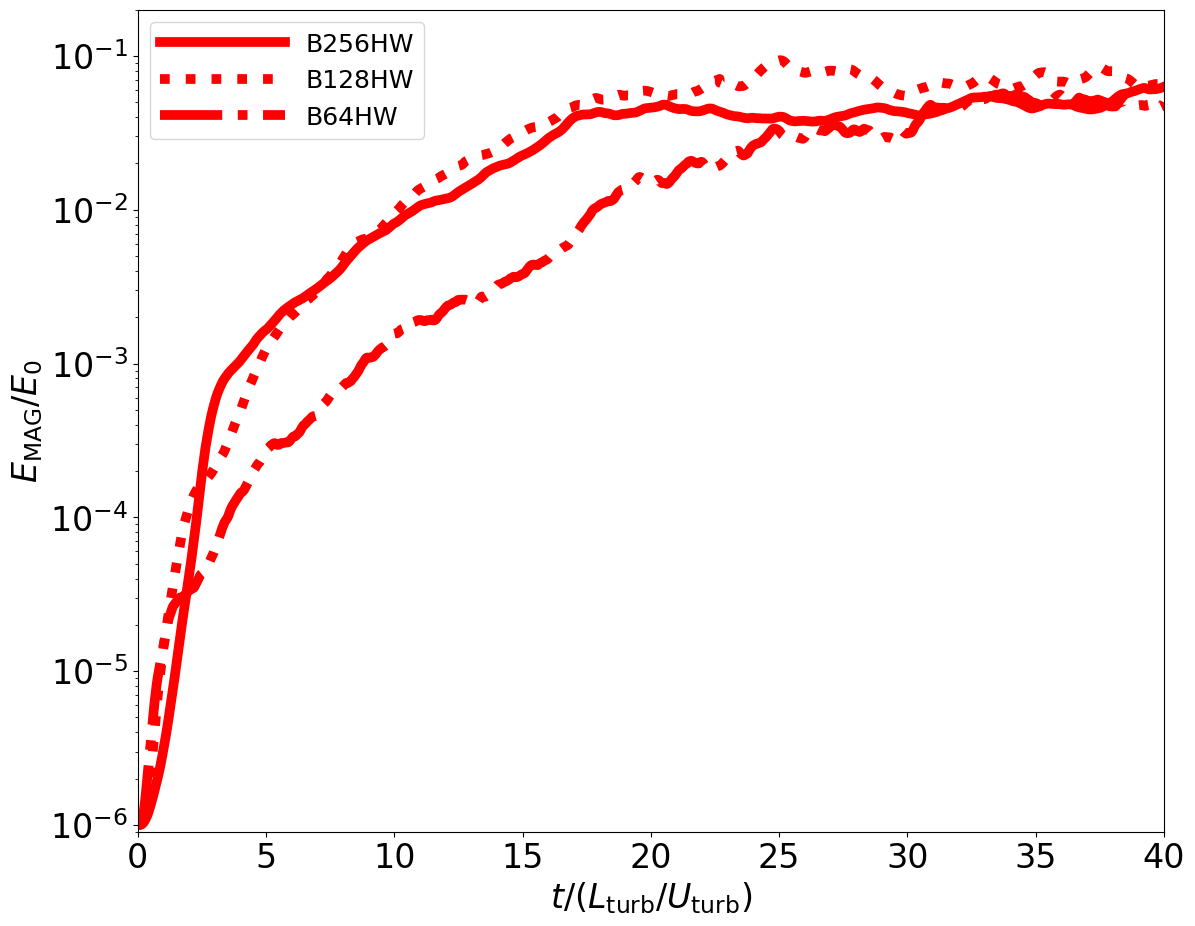}
    \caption{Magnetic energy density is shown for model A (purple curves in the top panel) and B (red curves in the bottom panel). The curves compare different resolutions $256^3$ (solid line), $128^3$ (dotted line), and $64^3$ (dash-dotted line). The reference energy density is defined as $E_{0} = \frac{1}{2} \rho_{0} U_{\rm turb}$.}
    \label{fig:compared_graphics_convergence_res}
\end{figure}

Figure \ref{fig:compared_graphics_convergence_res} compares the evolution of magnetic energy for three resolutions: \(64^3\), \(128^3\), and \(256^3\), focusing on the two main models, AXHW and BXHW, where X represents the resolution.  
In model A, the \(64^3\) resolution exhibits a shorter exponential and linear phase with a lower growth rate than the higher resolutions. The resolutions \(128^3\) and \(256^3\) display nearly identical exponential phases. The primary distinction lies in the linear phase: \(128^3\) is slightly slower, although both reach saturation at \(t \approx 15 (L_{\rm turb}/ U_{\rm turb})\). For model B, the \(64^3\) resolution also demonstrates a shorter exponential phase and a slower linear phase than the higher resolutions. In contrast, the \(128^3\) and \(256^3\) resolutions exhibit similar exponential phases and converge to saturation at \(t \approx 20 (L_{\rm turb}/ U_{\rm turb})\). Thus, for models A and B, the magnetic energy evolution at the \(128^3\) and \(256^3\) resolutions is very similar, suggesting convergence in this simulation aspect.

\begin{figure*}
    \centering
    \includegraphics[trim=260 138 200 130, clip, width=0.640\columnwidth]{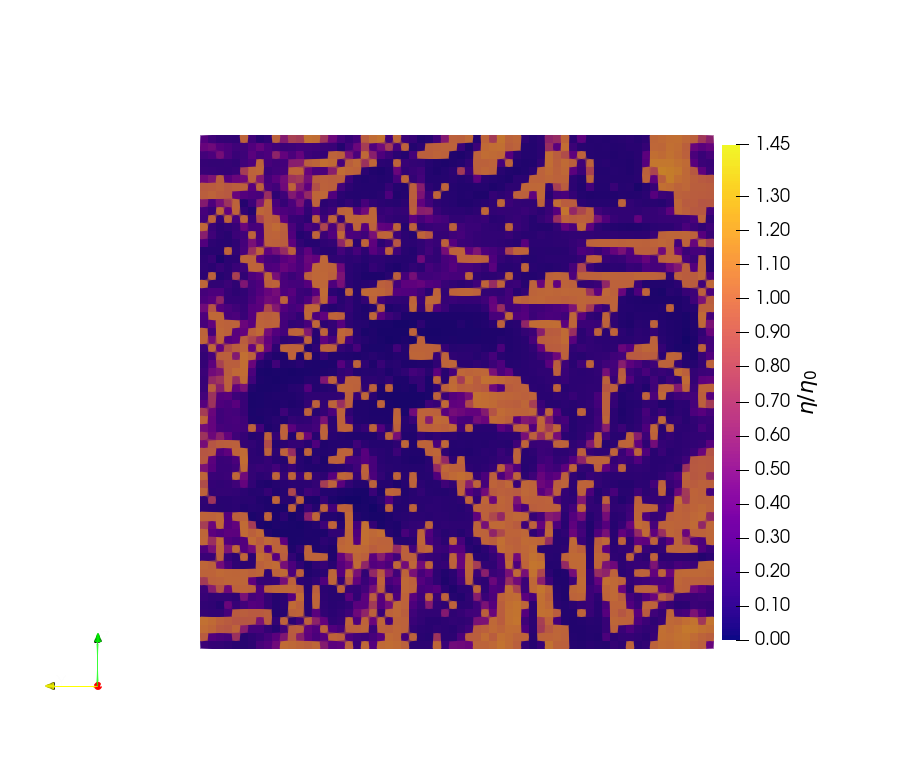}
    \includegraphics[trim=260 138 200 130, clip, width=0.640\columnwidth]{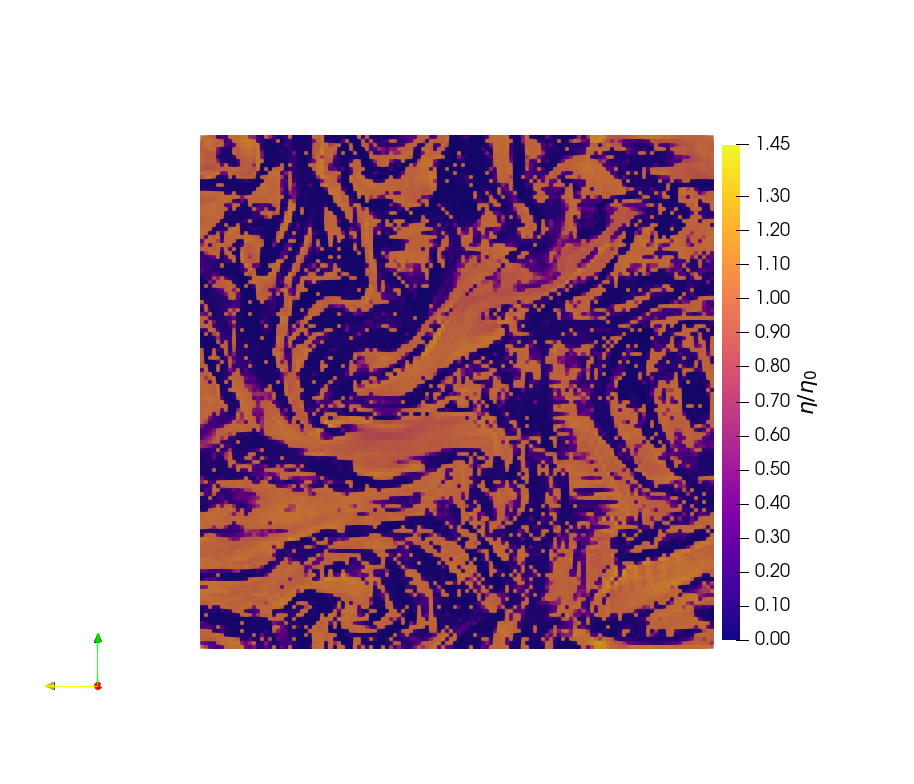}
    \includegraphics[trim=260 138 90 130, clip, width=0.795\columnwidth]{NUB_A256HW.png}
    \includegraphics[trim=260 138 200 130, clip, width=0.640\columnwidth]{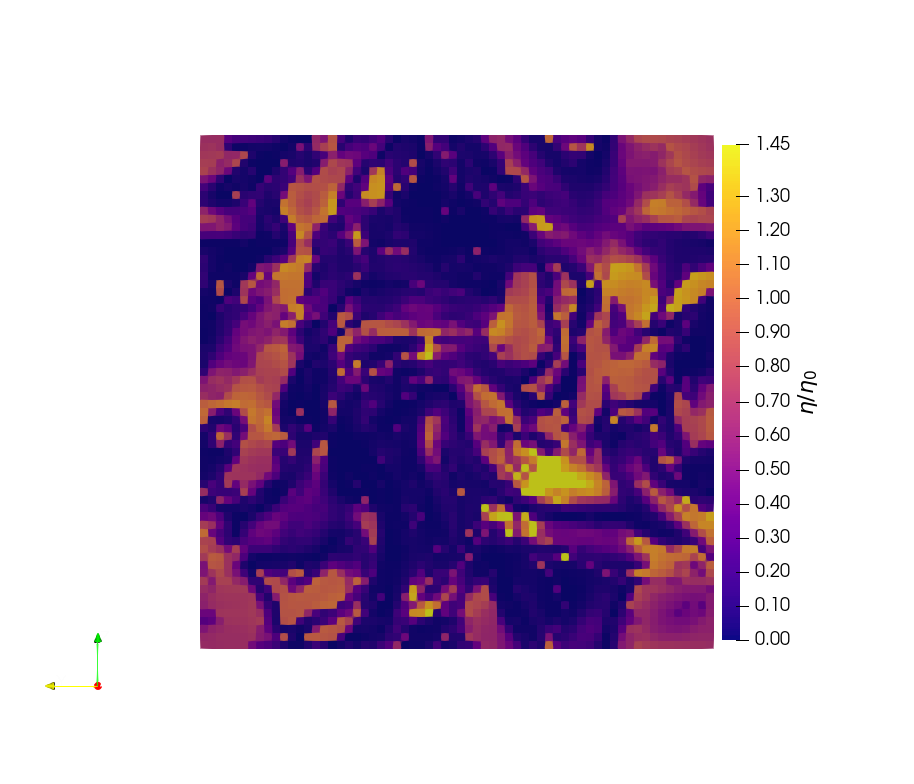}
    \includegraphics[trim=260 138 200 130, clip, width=0.640\columnwidth]{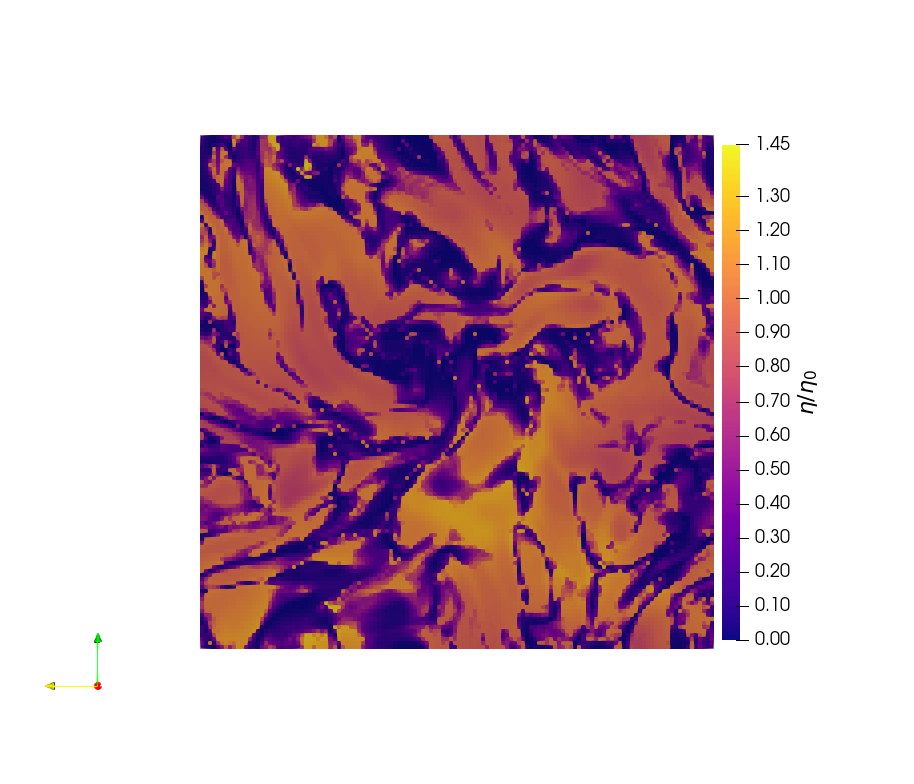}
    \includegraphics[trim=260 138 90 130, clip, width=0.795\columnwidth]{NUB_B256HW.png}
    \caption{Central XY distribution of the effective kinematic viscosity ($\eta/\eta_0$) for the models A064HW (top left), A128HW (top middle), A256HW (top right), and models B064HW (bottom left), B128HW (bottom middle), and B256HW (bottom right). The maps correspond to snapshots $t = 40 L_{\rm turb}/U_{\rm turb}$.}
    \label{fig:mapa_coeficiente_viscoso-AllRes}
\end{figure*}

Figure \ref{fig:mapa_coeficiente_viscoso-AllRes} compares the maps of the effective viscosity coefficient for runs AXHW and BXHW at resolutions \(64^3\), \(128^3\), and \(256^3\) in a central XY plane slice.  
In model A064HW, there is no obvious presence of the filamentary and fragmented structures observed at higher resolutions. Additionally, there is a visually larger portion of plasma with lower $\eta$ values. In the runs A128HW and A256HW, the filamentary pattern of \(\eta\) is evident at both resolutions. However, at \(256^3\), the darker structures are narrower and more fragmented compared to those at \(128^3\). Run B064HW also shows most of the plasma with lower \(\eta\) values at lower resolution, but this proportion decreases as the resolution increases. 
In contrast to model A, model B exhibits smoother structures with less stripping at any resolution. Overall, increasing the resolution for both models seems to result in more plasma returning to a stable state.

For a more detailed analysis of how numerical resolution affects the unstable fraction of the plasma and the mode evolution of viscosity, we refer to Appendix~\ref{apdx:resolution_effects}.


\section{Discussion}\label{sec:05}

In this study, we build upon previous CGL-MHD models, expanding the formulation of \citet{santos2014magnetic}. Using a new approach, we analyze the evolution of viscosity structure in a scenario of magnetic field amplification from initially weak fields with forced turbulence.  

The C1 model of \citet{santos2014magnetic} assumes instantaneous relaxation of pressure anisotropy (\(\nu_S = \infty\)), leading to dynamical evolution similar to a collisional non-viscous MHD model, with rapid magnetic energy growth and slightly lower saturation. In contrast, their C4 model adopts finite relaxation (\(\nu_S = 10^2\)), allowing moderate anisotropy development, which reduces the growth rate and results in lower magnetic energy saturation.
Although similar to the C4 model of \citet{santos2014magnetic}, our results are more consistent with the C1 model, where the relaxation rate of pressure anisotropy to the threshold set by the firehose and mirror instabilities is extremely rapid. The run A256HW saturates at \( t = 8 (L_{\rm turb}/U_{\rm turb}) \) at small scales, marking the transition from the exponential to the linear dynamo phase, and saturates at large scales after \( t = 10 (L_{\rm turb}/U_{\rm turb}) \), a behavior also observed in the C1 model of \citet{santos2014magnetic}.  

Figures \ref{fig:compared_graphics}, \ref{fig:evolution_moda_coefVisc_below01_percent}, and \ref{fig:coefficient_viscosity_evolucao_log_nub} highlight the intrinsic relationship between the growth of magnetic fields by the turbulent dynamo, the fraction of marginally unstable plasma, and the evolution of the effective viscosity coefficient. During the initial phase, there is a sharp reduction in the effective viscosity coefficient (Fig. \ref{fig:coefficient_viscosity_evolucao_log_nub}) alongside nearly 100\% of plasma in the unstable regime (Fig. \ref{fig:evolution_moda_coefVisc_below01_percent}). As the dynamo progresses, the fraction of unstable plasma gradually decreases, particularly in the runs A256HW and  A256HWnGs,
while the effective viscosity coefficient returns to values near its initial implementation. This transition marks the onset of the linear amplification phase, where global turbulent mechanisms sustain magnetic field growth. In the final phase, dynamo saturation is reached, stabilizing the magnetic field even under higher viscosity conditions over most of the plasma volume. These results highlight that the initial reduction in viscosity plays a critical role in dynamo efficiency, while subsequent evolution reflects a dynamic balance between turbulence and viscosity. This dynamical behavior underscores the importance of understanding how effective viscosity influences dynamo behavior at various stages, particularly in turbulent astrophysical environments.

The Braginskii-MHD model (model B in our analysis) also addresses magnetic field evolution in weakly collisional plasmas, assuming instantaneous relaxation in the viscous tensor to limit the pressure anisotropy at the firehose and mirror instability thresholds. In our simulations, following \citet{st2020fluctuation}, run B256HW employs hard-wall limiters similar to the L1 model described by the authors, with equivalent initial conditions.  In this model, pressure anisotropy is strictly confined within stability limits, resulting in dynamics similar to isotropic MHD. Our results are obtained in a significantly shorter time than those in \citet{st2020fluctuation}, likely due to methodological differences, such as the use of a compressible finite-volume code instead of the incompressible pseudospectral approach adopted in their study.
Although the dynamo is less efficient in this scenario than in CGL-MHD or collisional MHD without viscosity, it produces a final mean field comparable to MHD, consistent with those authors’ findings. Furthermore, we observe a shift in the peak of the magnetic power spectrum from \( k \approx 10 \times 2\pi/L \) during the exponential phase to \( k \approx 5 \times 2\pi/L \) (see Figures \ref{fig:power_spectrum} and \ref{fig:compared_graphics}). \footnote{This shift reflects the evolution of the spectrum, where the peak initially occurs at smaller scales and migrates to larger scales during the linear regime of the dynamo, before stabilizing (similar to the Kazantsev spectrum before the peak).} 
Despite methodological differences—using a compressible finite-volume code rather than the incompressible pseudospectral approach of the authors—our dynamo phases exhibit proportionally similar evolution. Although we did not replicate results without hard-wall limiters, we found similarities with our viscous MHD model, producing final values comparable to those reported in \citet{st2020fluctuation}. In particular, they analyzed in their U1 model, an unlimited Braginskii-MHD dynamo where pressure anisotropy evolves freely, generating sharp velocity gradients along the magnetic field direction. In our viscous MHD model, despite the absence of an inertial range, the amplification proceeds from small to large scales due to chaotic fluid motions. In contrast, \citet{st2020fluctuation} reports energy decay at small scales, consistent with a Kolmogorov spectrum. Despite methodological differences, the magnetic field growth curve in our simulations closely follows their U1 model.

Simulations using Braginskii or viscous MHD models provide valuable insights into physical processes in the intracluster medium but face limitations in fully capturing turbulent dynamo behavior and viscosity distribution. The Braginskii model, incorporating viscosity stress aligned with the magnetic field, more accurately reflects anisotropic effects induced by magnetic fields. In contrast, the viscous MHD model, assuming isotropic viscosity, simplifies momentum transport and neglects local variations tied to magnetic field topology. Nonetheless,  the Braginskii and viscous MHD models serve as essential starting points for understanding the interplay between viscosity and magnetic amplification in the ICM, especially on macroscopic scales.

Resolution dependency analyses reveal significant differences only at lower resolutions (\(64^3\)), while intermediate (\(128^3\)) and high (\(256^3\)) resolutions show consistent results. In model A, the exponential phase of magnetic energy amplification is shorter, and linear growth is slower at \(64^3\), with subtle differences between \(128^3\) and \(256^3\), indicating partial convergence. Similarly, in model B, evolution at \(128^3\) and \(256^3\) is nearly identical, both reaching saturation at equivalent times, whereas \(64^3\) exhibits greater instability and slower transitions. The spatial distributions of effective viscosity reflect these discrepancies, with \(64^3\) displaying less fragmented structures and a higher fraction of unstable plasma, while higher resolutions exhibit more detailed patterns and greater stabilization. These findings highlight that the low resolution typical of large-scale cosmological simulations can be inadequate for capturing the expected viscosity distribution.

\begin{figure}[ht]
    \centering
    \includegraphics[width=1.0\columnwidth]{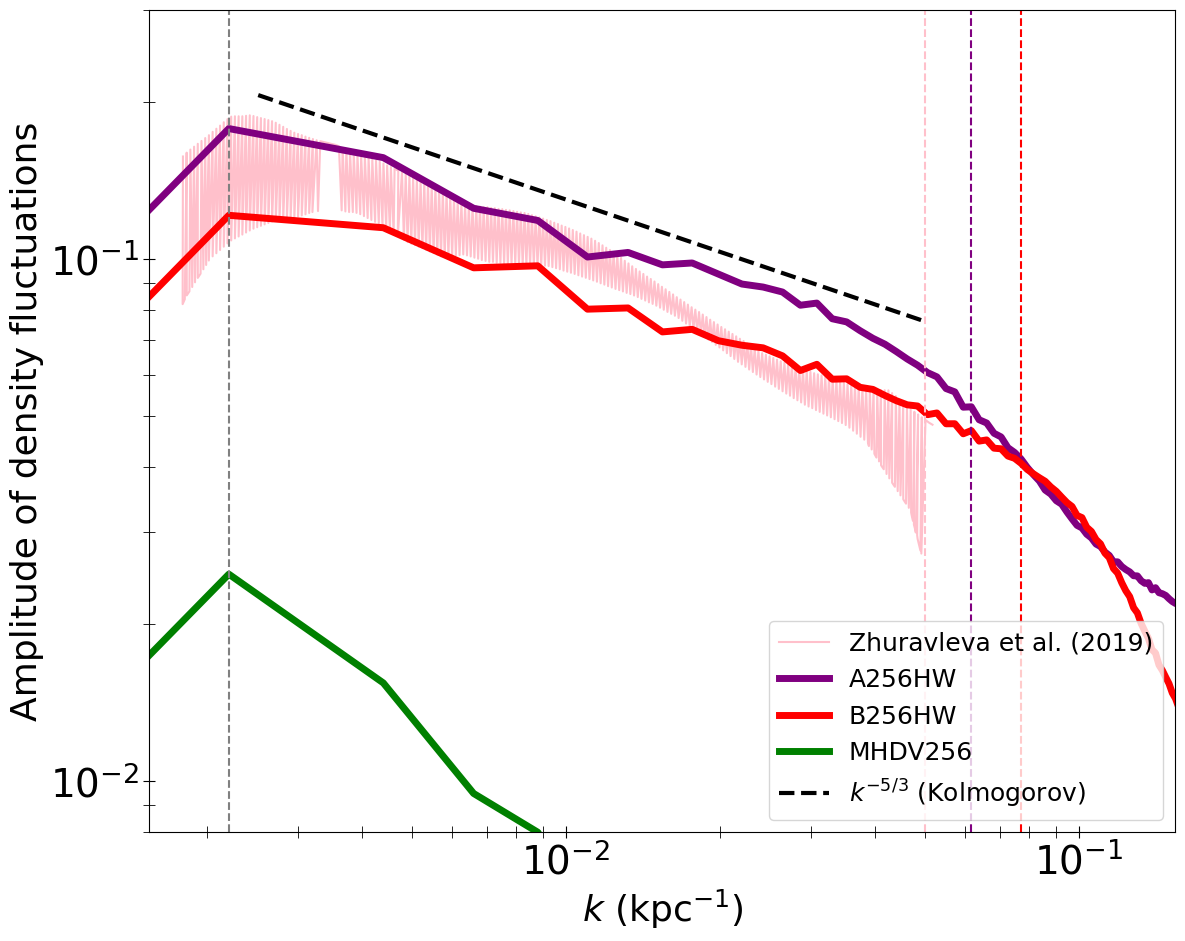}
    \caption{Comparison of the final stages of the turbulent velocity spectra for runs A256HW, B256HW, and MHDV256  from this work with the observed spectrum of the Coma cluster \citep{zhuravleva2019suppressed}. The consistency between the inertial range of the observed spectrum and model A especially reinforces the weakly collisional properties of galaxy clusters. The gray dashed line marks the onset of the inertial range for all models, the blue dashed line indicates its upper limit as observed by \citet{zhuravleva2019suppressed}, while the orange and magenta dashed lines represent the upper limits for models A and B, respectively.  The black dashed line represents the power law for the Kolmogorov spectrum ($\propto k^{-5/3}$). }
    \label{fig:zhuravleva_comparation}
\end{figure}

The spectrum of density fluctuations reported by \citet{zhuravleva2019suppressed} for the Coma cluster indicates significantly lower viscosity and smaller viscous scales than those predicted by Spitzer viscosity in a collisional fluid. Figure \ref{fig:zhuravleva_comparation} compares the spectra of the amplitude of density fluctuations obtained from our models A, B, and MHDV with the observed spectrum from the Coma cluster.

To derive the spectra from our models, we followed the normalization procedure described in Appendix~\ref{apdx:normalization_spectra}. 
Data adjustments in \citet{zhuravleva2019suppressed} assume a 700 kpc system, with wave numbers converted to physical units via \(k_{\text{physical}} = k_{\text{sim}} / L_{\text{physical}}\), where \(L_{\text{physical}} = 0.7\) Mpc. The corresponding amplitude spectrum of density fluctuations from our simulations is computed as $A_{3D} \propto \sqrt{2 k P_{\text{sim}}(k)/{3}}$ where \( P_{\text{sim}} \) is the simulated velocity power spectrum  (Figure \ref{fig:power_spectrum_final}, top panel). This procedure ensures consistency between our simulated spectra and the observationally inferred values (see Appendix~\ref{apdx:normalization_spectra} for more details).

In Figure \ref{fig:zhuravleva_comparation}, the inertial range of the weakly collisional models, particularly model A (extended CGL-MHD), spanning approximately \(2.2 \times 10^{-3}\) to \( \sim 6 \times 10^{-2}\), is comparable to the observed spectrum,  which extends from \(\sim 2.2 \times 10^{-3}\) to \(5 \times 10^{-2}\) kpc$^{-1}$, reinforcing the weakly collisional nature of the galaxy cluster plasmas. 
  
Notably, the observed spectrum does not show a clear change in slope at the tail, which would indicate the end of the inertial interval and a well-defined viscous scale. This is likely due to the limited sensitivity of the current X-ray instruments, which can hinder the detection of smaller structures within the cluster’s turbulent plasma. If this is the case, the actual viscous scale may be even smaller, as suggested by the weakly collisional models, particularly model A. Conversely, at the injection scale, the shape of the observed turbulent spectrum aligns well with the theoretical predictions, especially with model A. Meanwhile, the Braginskii-MHD (B) model more closely matches the observed spectrum at smaller scales. On the other hand,
the MHDV model does not present a well-defined inertial range. Overall, model A emerges as the best approximation for future high-resolution observations, although uncertainties remain regarding which model best represents current data.

Based on the complex structuring of the viscosity shown by our analysis, several questions regarding the physical processes in ICM could be explored in this context. For instance, the dissipation scales of different MHD waves evolve, directly influencing the propagation of cosmic rays through resonant interactions with these waves, the dynamics of shocks and galaxy wakes in this turbulent medium require further investigation to understand their role in shaping the magnetic field and plasma properties. Another critical issue is the evolution and structuring of viscosity under nonuniform forcing conditions, such as those driven by merger events or AGN jets in the central regions of a cluster, which introduce intermittent turbulence.

\section{Summary and Conclusions}\label{sec:06}

In this work, we study the evolution and structuring of viscosity in the intracluster medium (ICM) during turbulent magnetic field amplification. 
We developed periodic simulations of forced turbulence in a uniform and weak magnetic field to represent a portion of the ICM. In our main model, we employed 
the CGL-MHD equations for a fluid with an anisotropic pressure tensor, with anisotropy, evolved through the CGL closure. We extended these equations to introduce the effects of rapid anisotropy limitation caused by plasma instabilities, specifically mirror, and firehose, ensuring that the anisotropy does not exceed the threshold for these instabilities (``hard walls''). We studied the evolution of magnetic and velocity energy densities, magnetic field spectra, and the distribution of effective plasma viscosity, accounting for the collisionality introduced by the instabilities to maintain anisotropy within limits.

We compared the results of the extended CGL-MHD model with those from a Braginskii-MHD model, which assumes isotropic pressure and an explicit term for anisotropic viscosity (Braginskii viscosity). In this context, this model can be interpreted as a simplified representation of the stress tensor due to thermal pressure anisotropy, in the limit of small anisotropies. We also limited the action of this term to keep the anisotropy within the instability thresholds of the mirror and firehose. This second model has the advantage of being more easily implemented in simulations of the ICM. {\color{black}However, this simplicity comes at the cost of assuming a quasi-static evolution for the pressure anisotropy ($\mathrm{d}A/\mathrm{d}t \approx 0$), which can be relaxed by introducing a dynamic relaxation equation for the anisotropy. Such approaches allow the pressure anisotropy to evolve toward the Braginskii regime over time, bridging the gap between Braginskii-MHD and full CGL-MHD dynamics. These methods may be explored through hybrid kinetic-MHD models or reduced closures that interpolate between both regimes, and we plan to investigate them in future work.}

Furthermore, we compared these results with those from standard MHD simulations, which assume isotropic pressure and uniform explicit viscosity, commonly used in ICM studies, and have also considered an MHD model with no explicit viscosity.

The turbulence power injection rate was maintained at a similar level across all models. In the absence of a viscosity coefficient, this leads to a turbulence velocity amplitude comparable to the plasma sound speed, which was used to estimate the effective Reynolds number for all models. In our simulations, the collisional viscosity coefficient of ions (in the absence of instabilities) was adjusted to yield an initial Reynolds number of \(R_e \sim 30\).

Our main results can be summarized as follows:

\begin{itemize}

\item The magnetic energy in the extended CGL-MHD model (model A in our analysis) grows to the same levels as the MHD model without viscosity, independent of the seed field. In the Braginskii-MHD model model B in our analysis), the magnetic field growth rate during the linear phase shows a small difference, being slower.

\item In the initial phase (\(t < 5 L_{\rm turb} / U_{\rm turb}\)) of the small-scale dynamo, the effective  viscosity in the extended CGL-MHD model peaks around 
\( 3.4 \times 10^{-4} L_{\rm turb} U_{\rm turb}\), corresponding to an average $<\eta>/\eta_0 \simeq  3 \times 10^{-2}$, due to the presence of plasma instabilities in most of the domain. As the magnetic field amplifies (and the \(\beta\) parameter decreases), the viscosity distribution becomes bimodal peaking around: (i) the initial collisional value $1.67 \times 10^{-2} L_{\rm turb} U_{\rm turb}$, and (ii) $1 \times 10^{-4} L_{\rm turb} U_{\rm turb}$ in the volume fraction where the instabilities are excited. In the saturated phase of the dynamo, about 60\% of the plasma volume remains stable for the mirror and firehose instabilities, with viscosity attaining the collisional value. At the same time, the rest of the system exhibits the low viscosity values of the unstable zone $1 \times 10^{-4} L_{\rm turb} U_{\rm turb}$, and an average $<\eta>/\eta_0 \simeq  1$.

\item The marginally unstable plasma seems to concentrate in laminar regions of high vorticity/shear. In the Braginskii-MHD model, the viscosity distribution across the domain is smoother.

\item In the Braginskii-MHD model, during the initial phase of the dynamo (\(t < 5 L_{\rm turb} / U_{\rm turb}\)), the viscosity regulated by instabilities is lower compared to that in the CGL-MHD model, reaching approximately \(5 \times 10^{-5} L_{\rm turb} U_{\rm turb}\), and $<\eta>/\eta_0 = 10^{-4}$. Compared to the CGL-MHD model, a larger volume of unstable plasma remains in the saturated phase, with about 50\%. 

\item In the saturated phase of the dynamo, the spatial scale at which viscous dissipation becomes significant in the velocity spectrum of the extended CGL-MHD model is larger than in the MHD model without viscosity. In the Braginskii-MHD model, this scale is smaller than in the CGL-MHD model. In the MHD model with explicit viscosity similar to the average viscosity in the saturated phase of the CGL-MHD model, turbulence does not exhibit an inertial range.

\item The magnetic field spectrum in the extended CGL-MHD and Braginskii-MHD models is very similar to the MHD model without viscosity in the saturated phase. The magnetic field spectrum in the MHD simulation with collisional viscosity similar to the average viscosity in the saturated phase of the CGL-MHD model, is also similar to the previous ones but with a lower amplitude.

\item The amplitude spectra of density fluctuations obtained from our extended CGL-MHD and Braginskii-MHD models are consistent with observations of the Coma cluster \citep{zhuravleva2019suppressed}, which extend to much smaller scales than those predicted for a collisional viscous plasma. This further supports the weakly collisional nature of galaxy cluster plasmas.

\end{itemize} 

{\color{black} Although our simulations adopt an initial homogeneous and non-stratified medium, it is important to note that real galaxy clusters exhibit stratification, with radial gradients in density, temperature, and magnetic field strength. These gradients would lead to local variations in plasma beta, turbulence intensity, and the thresholds for instability onset, potentially leading to layered structures in effective viscosity and spatially dependent dynamo efficiency. Although our approach focuses on average conditions to isolate the role of anisotropic viscosity and plasma instabilities, we expect the main conclusions, such as the development and saturation of pressure anisotropy and the comparative behavior of different viscosity models, to remain qualitatively robust. Future work, including stratification, will allow spatially resolved predictions and closer comparisons with observational data.}

{\color{black} A final note worth highlighting is that our use of continuous energy injection likely overestimates the magnetic field strength compared to real systems, where energy is injected intermittently via mergers. Under realistic conditions, the magnetic field may decay or never reach saturation between such events. Moreover, the adoption of a simplified isothermal equation of state may have influenced our results by neglecting local temperature variations. In future studies, we plan to address this by modeling non-uniform energy injection. }

\section*{Data availability}
The simulated data generated during this study are available at the request of the authors.

\section*{Acknowledgements} SAF acknowledges support from the Brazilian Funding Agency FAPESP (scholarships 2019/04992-4  and 2023/02753-8). EMdGDP also acknowledges support from FAPESP (grant 2021/02120-0) and the Brazilian Agency CNPq (grant 308643/2017-8).  The simulations presented in this work were performed in the cluster of the Group of Plasmas and High-Energy Astrophysics (GAPAE), acquired with support from FAPESP (grants 2013/10559-5 and 2021/02120-0), and the SYRTARI server acquired with support from FAPESP (grant 2019/05757-9).

\newpage

\appendix

{\color{black} 
\section{Time Evolution of Volume-Averaged Plasma Beta}
\label{sec:appendix_beta}

Figure~\ref{fig:beta_evolution} shows the time evolution of the volume-averaged plasma beta ($\langle \beta \rangle$) for all simulations. As turbulence amplifies the magnetic field, $\langle \beta \rangle$ decreases and reaches saturation levels that depend on the viscosity model. 

\begin{figure}[h]
\centering
\includegraphics[width=0.6\linewidth]{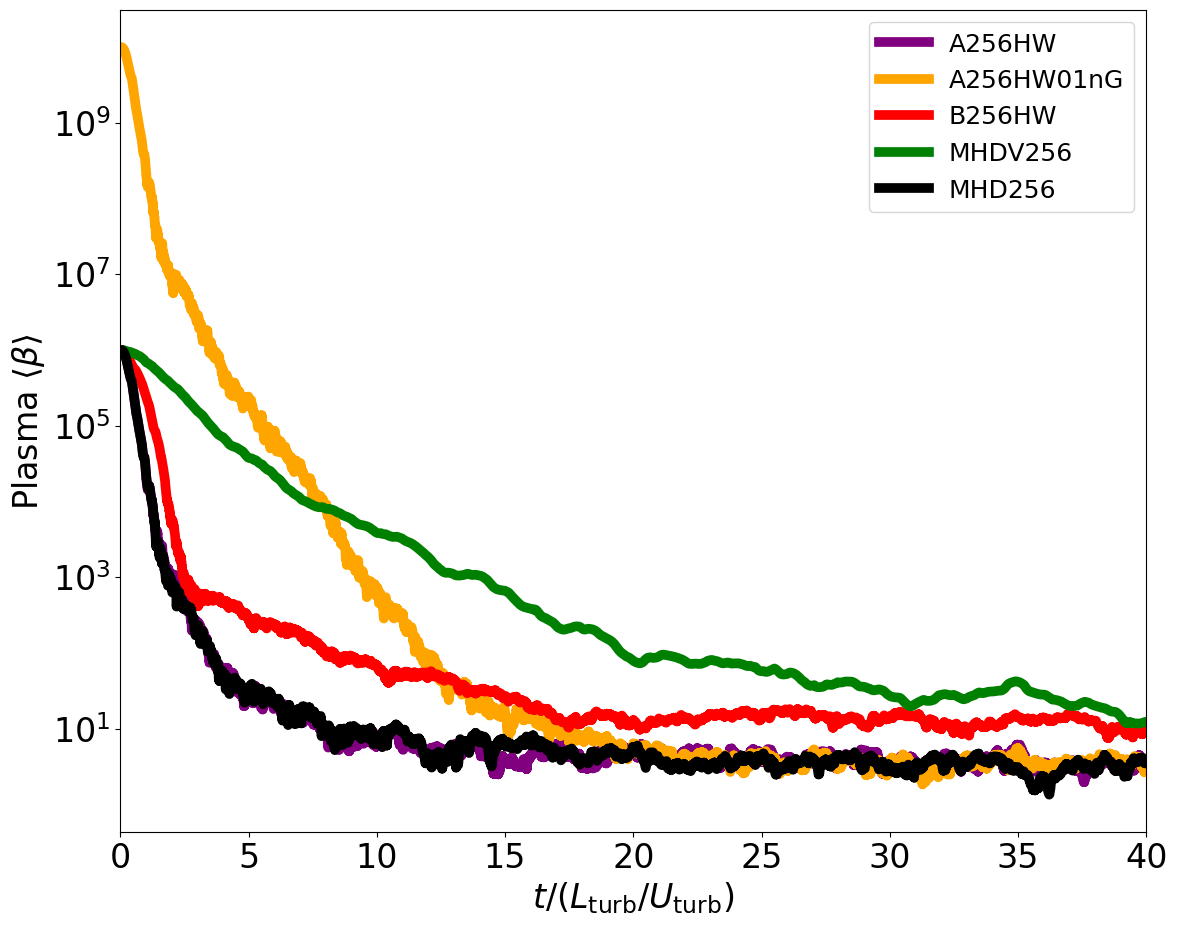}
\caption{
{\color{black} 
Time evolution of the volume-averaged plasma beta ($\langle \beta \rangle$) for all simulations: A256HW, A256HWnG, B256HW, MHDV256, and MHD256. Despite high initial values ($\beta_0 \sim 10^6$--$10^{10}$), turbulent amplification leads to saturation at $\langle \beta \rangle \sim 20$--$200$.}}
\label{fig:beta_evolution}
\end{figure}}

\section{Evolution of the Average of the Effective Viscosity Coefficients}
\label{apdx:viscosity_evolution}

\begin{figure}
    \centering
    \includegraphics[width=0.6\columnwidth]{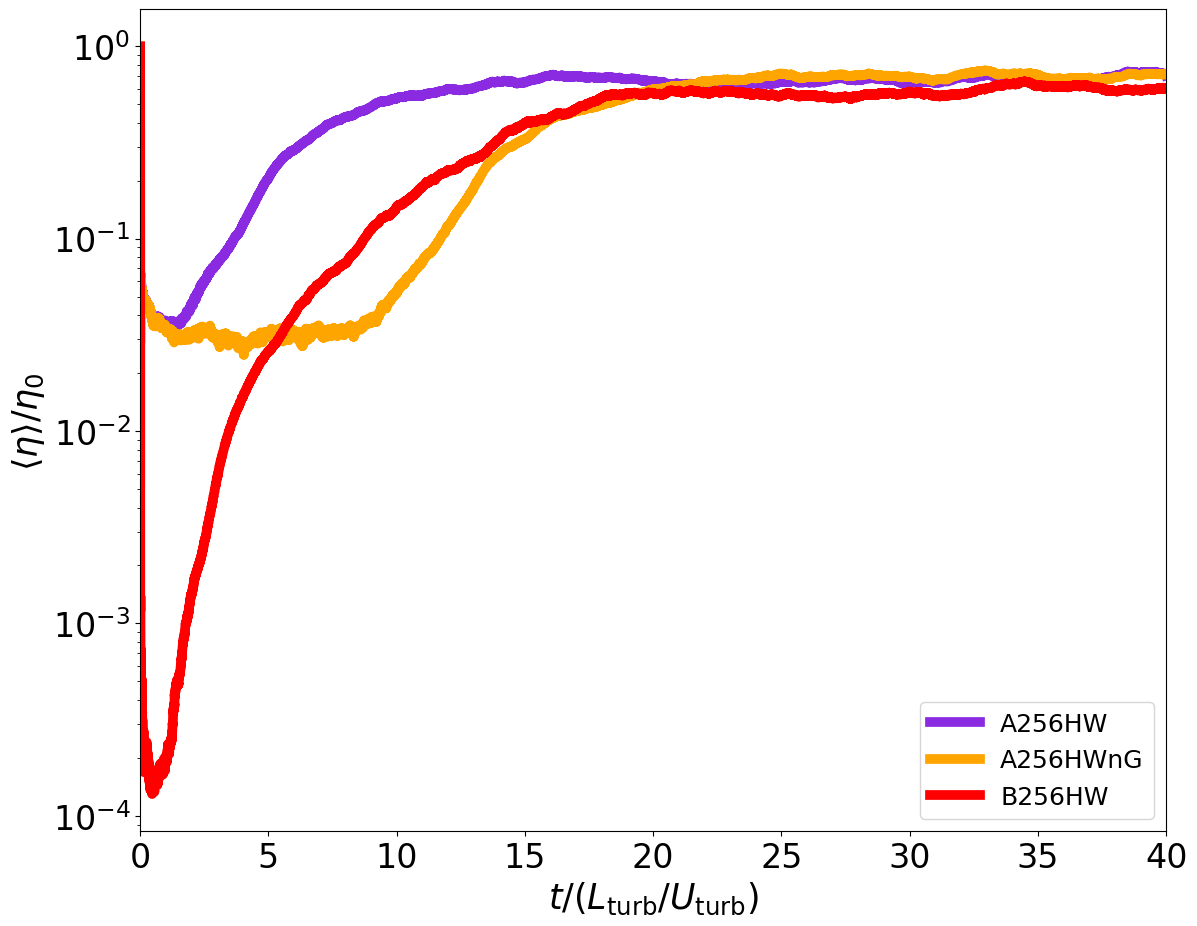}
    \caption{Evolution of the $average$ effective viscosity coefficients for the entire system for models  A256HW (purple curve), A256HWnG (orange curve), and B256HW (red curve).}
    \label{fig:coefficient_viscosity_evolucao_log_nub}
\end{figure}

Figure~\ref{fig:coefficient_viscosity_evolucao_log_nub} presents the time evolution of the average of the effective viscosity coefficients for the entire system in models A256HW, A256HWnG, and B256HW.

For the A256HW run, the average initially decreases to values close to \(3 \times 10^{-2}\), followed by a rapid increase until stabilizing around \(6 \times 10^{-1}\) at \(t \sim 15 (L_{\rm turb}/U_{\rm turb})\). The A256HWnG model exhibits a similar initial behavior but reaches equilibrium later, stabilizing at approximately \(6 \times 10^{-1}\) only at \(t \sim 20 (L_{\rm turb}/U_{\rm turb})\). This delay aligns with the evolution of dynamo-driven magnetic field amplification, as seen in Figure~\ref{fig:compared_graphics}.

In contrast, the B256HW model experiences a much sharper initial drop, reaching values near \(10^{-4}\), before rapidly growing and stabilizing around \(6 \times 10^{-1}\) at \(t \sim 20 (L_{\rm turb}/U_{\rm turb})\), consistent with its magnetic field evolution in Figure~\ref{fig:compared_graphics}.

The sharp initial decrease in effective viscosity coefficients is due to the injection of turbulence, which initially reduces all viscosity values. This trend aligns with the behavior of the mode in Figure~\ref{fig:combined_moda_viscosity} and with the evolution of the viscosity coefficient histogram in Figure~\ref{fig:evolution_histogram_coefVisc}, where the distribution initially shifts toward lower values.

\section{Resolution Effects on Unstable Plasma Fraction and the Evolution of Viscosity}
\label{apdx:resolution_effects}

\begin{figure}
    \centering
    \includegraphics[width=0.6\columnwidth]{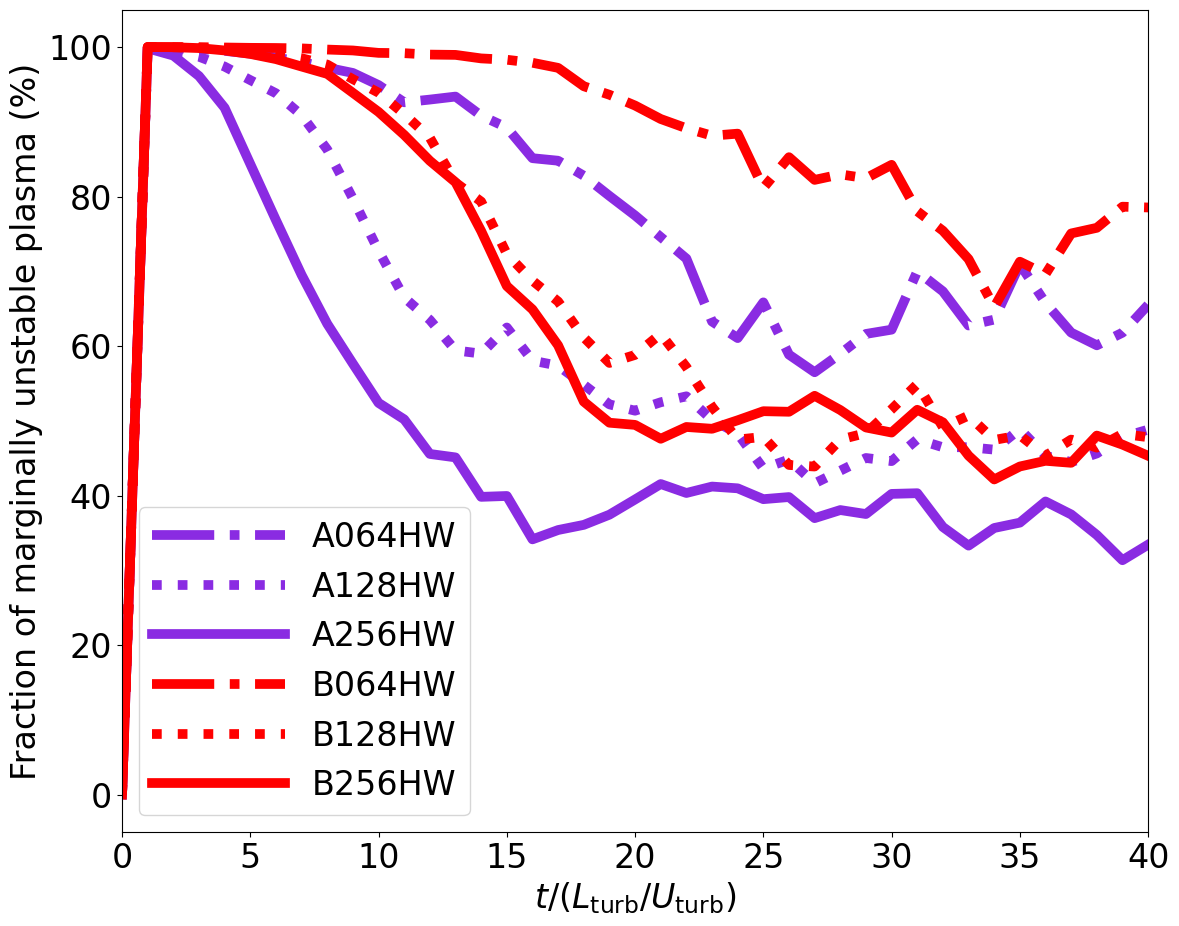}
    \caption{Evolution of the fraction of plasma at the instabilities threshold for models A256HW (purple solid curve), A128HW (purple dotted curve), A064HW (purple dashdot curve), B256HW (red solid curve), B128HW (red dotted curve), and B064HW (red dashdot curve).}
    \label{fig:evolution_moda_coefVisc_below01_percent_allRes}
\end{figure}

Figure \ref{fig:evolution_moda_coefVisc_below01_percent_allRes} compares the evolution of the unstable plasma fraction (in \(\%\)) for resolutions \(64^3\), \(128^3\), and \(256^3\) in models A and B. In model A, the unstable fraction stabilizes at ~70\% for \(64^3\), decreases to ~50\% for \(128^3\), and reaches ~40\% for \(256^3\). In model B, the \(64^3\) resolution stabilizes at ~80\%, while \(128^3\) and \(256^3\) converge to ~50\%.

\begin{figure}
    \centering
    \includegraphics[width=0.48\columnwidth]{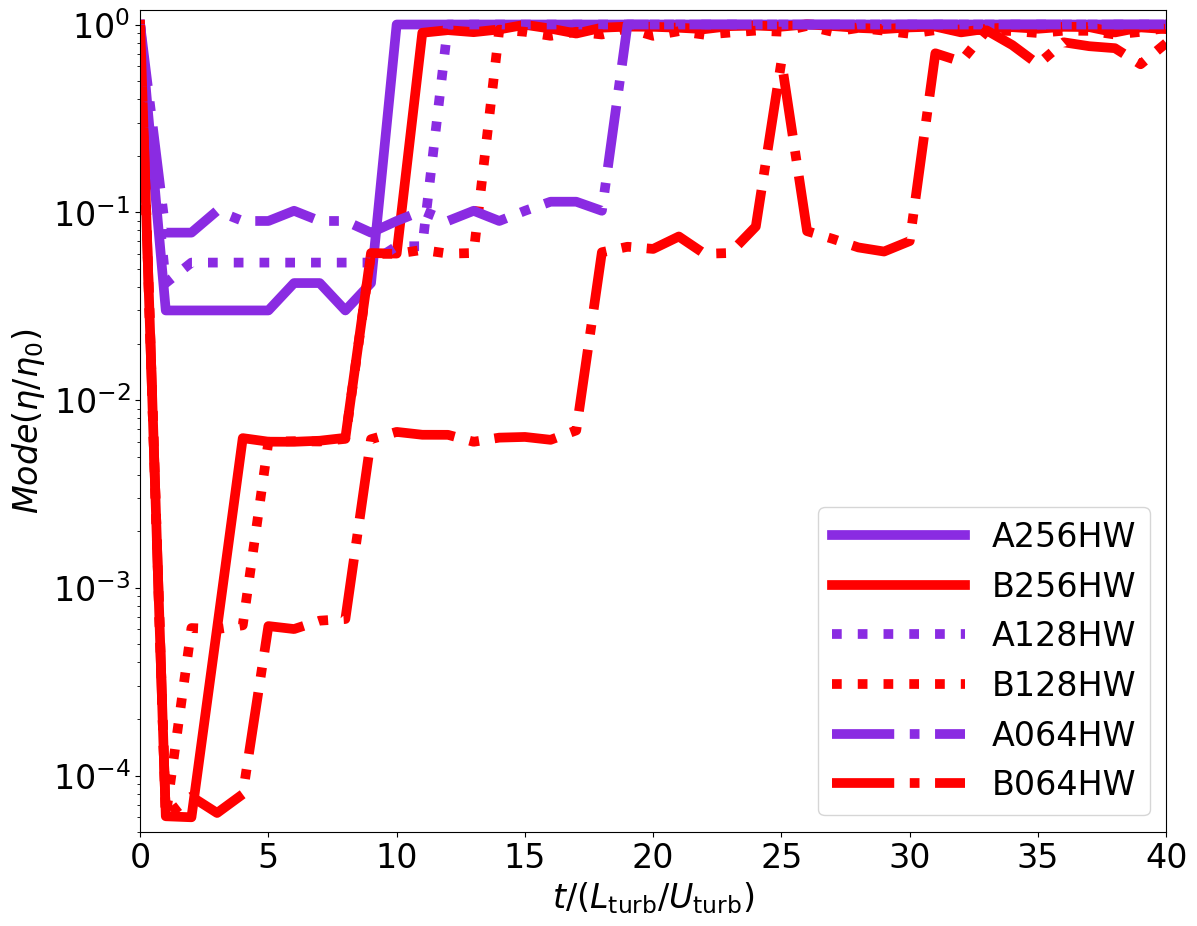}
    \includegraphics[width=0.48\columnwidth]{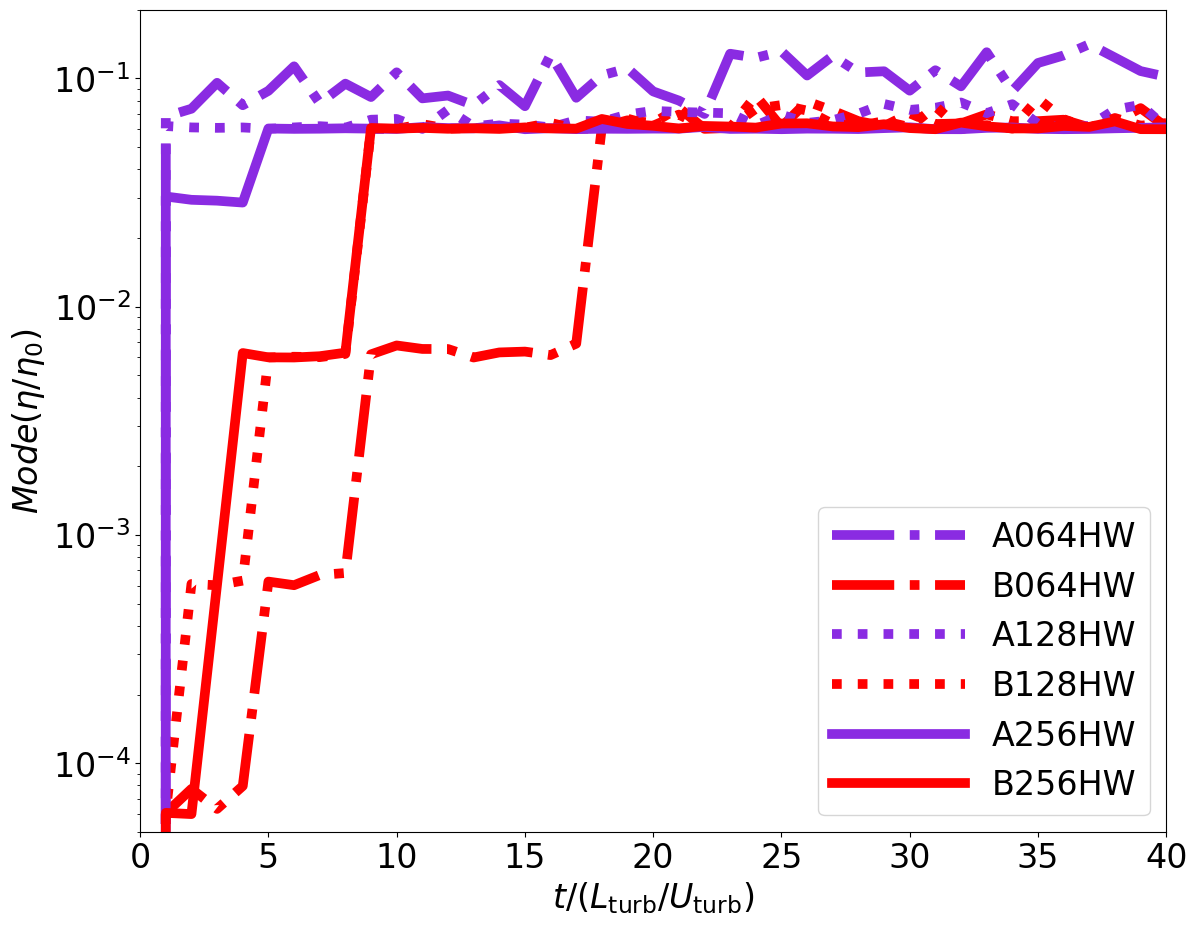}
    \caption{Evolution of the mode of the effective viscosity coefficients for models A256HW (purple solid curve), A128HW (purple dotted curve), A064HW (purple dashdot curve), B256HW (red solid curve), B128HW (red dotted curve), and B064HW (red dashdot curve). The upper panel shows the evolution of the entire plasma volume, while the lower panel considers only the fraction of marginally unstable plasma.}
    \label{fig:combined_moda_viscosity_resolution}
\end{figure}

Figure \ref{fig:combined_moda_viscosity_resolution} shows the evolution of the mode of \(\eta_{\rm eff} / \eta_0\) for the entire plasma (top panel) and the unstable fraction (bottom panel) in models A and B at resolutions \(64^3\), \(128^3\), and \(256^3\). In both models, the \(128^3\) and \(256^3\) resolutions exhibit similar trajectories, with \(Mode(\eta / \eta_0) \approx 1\) for the global plasma and \(7 \times 10^{-2}\) for the marginally unstable fraction. However, the \(64^3\) resolution shows stronger fluctuations, accelerated growth in the marginally unstable fraction for model A, and slower, irregular growth for model B. 

\section{Normalization and Unit Conversion for Spectral Comparison}
\label{apdx:normalization_spectra}

The following steps outline the procedure used to compare our simulated models with the observationally derived spectrum of the Coma cluster \citep{zhuravleva2019suppressed}.

According to \citet{zhuravleva2019suppressed}, the observed spectrum is scaled assuming a characteristic system size of 700 kpc, requiring the conversion of simulated wavenumbers into physical units via  

\begin{equation}
    k_{\text{physical}} = \frac{k_{\text{sim}}}{L_{\text{physical}}},
\end{equation}
where \( L_{\text{physical}} = 0.7 \) Mpc. 

At a given spatial scale \( l \), characterized by a wave number \( k = 1/l \), the observed amplitude of velocity and density fluctuations follows an approximately linear relation, given by \citep{zhuravleva2019suppressed}:

\begin{equation}\label{eq:A3d}
  A_{3D} =  \sqrt{4\pi k^3 P(k)} \propto \frac{V_{1k}}{c_s},
\end{equation}
where \( P(k) \) is the power spectrum of density fluctuations, \( c_s \) is the isothermal sound speed, and \( V_{1k} \) is the velocity component along one direction. We note that \citet{zhuravleva2019suppressed} adopted a sound speed $1.4 \times 10^8$ cm/s, which is similar to the value adopted in this work (Section  \ref{sec:setup}).   The relation between the velocity amplitude and the kinetic energy density spectrum \( E(k) \) is given by

\begin{equation}\label{eq:V1k}
    V_{1k}^2 = \frac{2}{3} k E(k).
\end{equation}

In order to compute the density fluctuation amplitude spectrum from our simulation data, we derive the power spectrum of density fluctuations, \( P(k) \), using Eqs.~\ref{eq:A3d} and \ref{eq:V1k}, leading to

\begin{equation}
    P(k) \propto \frac{E(k)}{6 \pi k^2 c_s^2}.
\end{equation}

Since in our simulations the velocity field spectrum is already normalized by the sound speed, the dimensionless power spectrum $P_{\text{sim}}$ can be expressed as 

\begin{equation}
    P_{\text{sim}}(k) = \frac{E(k)}{c_s^2}.
\end{equation}

Thus, the three-dimensional fluctuation amplitude spectrum follows.

\begin{equation}
    A_{3D} \propto \sqrt{\frac{2 k P_{\text{sim}}(k)}{3}}.
\end{equation}

We note that the constant of proportionality in this relation depends on the details of how the power spectrum is built. In \citet{zhuravleva2019suppressed}, this constant, or else the exact relation between the amplitude of the observed and simulated spectra, is not explicitly provided. Therefore, to match the amplitude of the spectrum obtained from the observations to our simulated spectra, we multiplied the latter by a factor $\sim 9 \sqrt{2\pi}$. {\color{black}The factor of $2\pi$ was included based on the definition of wavenumber adopted in \citet{zhuravleva2019suppressed}, where $k = 1/l$ (i.e., without the usual $2\pi$ factor), implying a conversion compared to our Fourier-transformed quantities. An additional factor of 9 was introduced to shift the simulated amplitude spectrum upward, 
making it easier to compare key quantities such as the shape and viscous scale.}


\vspace{5mm}

\bibliography{sample631}{}
\bibliographystyle{aasjournal}

\end{document}